\documentclass[conference]{IEEEtran}
\IEEEoverridecommandlockouts

\usepackage{cite}
\usepackage{amsmath,amssymb,amsfonts}
\usepackage{graphicx}
\usepackage{textcomp}
\usepackage{xcolor}
\def\BibTeX{{\rm B\kern-.05em{\sc i\kern-.025em b}\kern-.08em
    T\kern-.1667em\lower.7ex\hbox{E}\kern-.125emX}}

\usepackage{xcolor}
\usepackage{subfig}
\usepackage{multirow}
\usepackage[normalem]{ulem}
\usepackage{fancyhdr}
\usepackage{amsmath}
\usepackage{xspace}
\usepackage{soul}
\usepackage{placeins}

\usepackage{url}
\usepackage{etex}
\usepackage{etoolbox}
\usepackage{xspace}
\usepackage{amsmath}
\usepackage{xcolor}
\usepackage{nicefrac}
\usepackage{changepage}
\usepackage{array,booktabs,multirow}
\usepackage{afterpage}
\usepackage{pbox}
\RequirePackage{enumerate}
\RequirePackage{paralist}
\RequirePackage{enumitem}
\usepackage{paralist}
\usepackage{mdwlist}

\usepackage{hhline}
\usepackage{lipsum} 
\usepackage{adjustbox}
\usepackage{tabu}
\usepackage{makecell}
\usepackage{color, colortbl}
\usepackage{amsmath,amssymb,latexsym}
\usepackage{graphicx,dblfloatfix}
\usepackage{relsize}
\usepackage{multicol}

\usepackage{dblfloatfix}
\usepackage{blindtext}

\usepackage{amsfonts}

\usepackage{pifont}

\usepackage{algorithm}
\usepackage[noend]{algpseudocode}

\floatevery{algorithm}{\setlength\hsize{3.7cm}}

\algnewcommand{\LineComment}[1]{\State \(\triangleright\) #1}     
\makeatletter
\renewcommand{\ALG@beginalgorithmic}{\small}
\makeatother                        
                                                                                          
\algblockdefx{FORP}{ENDFP}[1]%
{\textbf{for all }#1 \textbf{do in parallel}}%
{\textbf{end for}}

\algblockdefx{FORB}{ENDFB}[1]%
{\textbf{for }#1 \textbf{do in backward}}%
{\textbf{end for}}

\makeatletter                                                                             
\newcommand{\algorithmfootnote}[2][\footnotesize]{%
    \let\old@algocf@finish\@algocf@finish
    \def\@algocf@finish{\old@algocf@finish
        \leavevmode\rlap{\begin{minipage}{\linewidth}                                     
                #1#2                                                                      
        \end{minipage}}%
    }%
}

\makeatletter                                                                             
\newcommand{\thickhline}{%
    \noalign {\ifnum 0=`}\fi \hrule height 1.2pt                                          
    \futurelet \reserved@a \@xhline                                                       
}                                                                                         
\makeatletter                                                                             
\newcommand{\midhline}{%
    \noalign {\ifnum 0=`}\fi \hrule height 1pt                                            
    \futurelet \reserved@a \@xhline                                                       
}    

\newcommand{\ceil}[1]{\lceil #1 \rceil}

\newcommand{\abbrfnt}{}
\newcommand{\abbrev}[1]{{\mbox{\abbrfnt{#1}}}\xspace}
\newcommand{\sect}[1]{Section.~\ref{#1}\xspace}

\newcommand{\eq}[1]{Eq.~\ref{#1}\xspace}

\makeatletter
\renewcommand{\fnum@table}{Tab. \thetable}
\makeatother
\newcommand{\tab}[1]{Tab.~\ref{#1}\xspace}

\makeatletter
\renewcommand{\fnum@figure}{Fig. \thefigure}
\makeatother
\newcommand{\fig}[1]{Fig.~\ref{#1}\xspace}

\makeatletter
\def\SOUL@hlpreamble{%
    \setul{0ex}{2ex}
    \let\SOUL@stcolor\SOUL@hlcolor
    \SOUL@stpreamble
}
\makeatother

\makeatletter  
\def\@eqnnum{{\normalfont\normalcolor[\theequation]}}  
\makeatother

\newcommand{\mdel}[1]{}

\newcommand{\FIXME}[1]{}

\newcommand{\del}{\abbrev{DeLTA}}
\newcommand{\conv}{\abbrev{conv}}
\newcommand{\scratch}[1]{}

\begin{document}                                                                          
\title{\huge DeLTA: GPU Performance Model for \underline{De}ep \underline{L}earning Applications with In-depth Memory System \underline{T}raffic \underline{A}nalysis}

\author{                                                                                  
    \font\athorFont=cmr12 at 11pt                                                         
    {\athorFont                                                                           
        Sangkug Lym\IEEEauthorrefmark{2} \hspace{0.6em}                                   
        Donghyuk Lee\IEEEauthorrefmark{3} \hspace{0.6em}                                    
        Mike O'Connor\IEEEauthorrefmark{2}\IEEEauthorrefmark{3} \hspace{0.6em}                                  
        Niladrish Chatterjee\IEEEauthorrefmark{3} \hspace{0.6em}                                
        Mattan Erez\IEEEauthorrefmark{2}                                                  
    }                                                                                     
    \vspace{8pt}\\                                                                        
    \font\athorFont=cmr12 at 11pt                                                         
    {\athorFont                                                                           
        \IEEEauthorrefmark{2}The University of Texas at Austin \hspace{2em}               
        \IEEEauthorrefmark{3}NVIDIA
    }                                                                                     
    \vspace{4pt}\\                                                                        
    \font\athorFont=cmr12 at 8pt                                                          
    {                                                                                     
        \athorFont                                                                        
        \IEEEauthorrefmark{2}{\textit{\small\{sklym, mattan.erez\}@utexas.edu}} \hspace{2em} 
        \IEEEauthorrefmark{3}{\textit{\small \{donghyukl, moconnor, nchatterjee\}@nvidia.com}}
    }                                                                                     
    \vspace{7pt}\\                                                                        
}

%


\maketitle
\begin{abstract}
Training convolutional neural networks (CNNs) requires intense compute throughput and high memory bandwidth.
Especially, convolution layers account for the majority of execution time of CNN training, and GPUs are commonly used to accelerate these layer workloads.
GPU design optimization for efficient CNN training acceleration requires the accurate modeling of how their performance improves when computing and memory resources are increased. 
We present \del, the first analytical model that accurately estimates the traffic at each GPU memory hierarchy level, while accounting for the complex reuse patterns of a parallel convolution algorithm.
We demonstrate that our model is both accurate and robust for different CNNs and GPU architectures. 
We then show how this model can be used to carefully balance the scaling of different GPU resources for efficient CNN performance improvement.
\end{abstract}

\begin{IEEEkeywords}
GPU, memory system, deep learning, CNN
\end{IEEEkeywords}

\vspace*{-3mm}
\section{introduction}
\label{sec:intro}

Convolutional neural networks (CNNs) are the state of the art for various vision applications~\cite{krizhevsky2012imagenet,ren2015faster,redmon2016you,wang2017fast}. The computation required for CNNs is a good match for GPUs~\cite{krizhevsky2012imagenet,simonyan2014very}. The increasing demand for CNN computation is driving GPU arithmetic performance, which has been increasing at higher than its historical rate~\cite{volta2017whitepaper}. However, based on our analysis, compared to the rapid GPU compute throughput increase of 32X for the past 9 years, its memory system bandwidth has improved by only 13X. This memory wall problem can bottleneck the performance of even the arithmetically-intensive CNNs, making performance scaling difficult.

It is therefore imperative to balance both arithmetic and memory performance in architecting a future GPU for efficient CNN performance scaling. This optimization benefits from analytical modeling, which can quickly provide insight and narrow the design space before slower and more resource-consuming modeling is used (e.g., simulators~\cite{aamodt2012gpgpu}). Analytical models also aide in the optimization of software for efficient HW resource utilization~\cite{zhou2017performance,lai2013performance}. Prior work on modeling CNN performance on GPUs focuses on modeling arithmetic performance with only simplistic and naive modeling of memory~\cite{hong2009analytical,lai2013throughput,lai2013performance,zhou2017performance}. While these models are accurate when performance is bound by arithmetic throughput, the growing memory wall has shifted performance bottleneck to the memory system in many cases. 

We present \del---the first analytical GPU model for CNNs that accurately models both arithmetic performance and traffic in the memory hierarchy, and that can therefore be used to identify performance bottlenecks and explore the optimization space for future designs.

Our GPU CNN performance model is both the first to model cache traffic in the GPU and the first to handle the most-commonly used algorithm for convolution (\conv) and fully-connected (FC) layers in GPUs---\emph{im2col} (image-to-column)~\cite{chetlur2014cudnn}. While a \conv layer can be computed by directly implementing the convolution operator, casting the convolution problem as a general matrix-matrix multiplication (GEMM) is both simpler and more efficient on a GPU. Our model is unique in that it accounts for the complex access locality that exists in im2col as the algorithm replicates and reorders input matrices to expose the parallelism needed for effective GPU acceleration. We demonstrate that only \del is accurate enough to be use for architectural exploration, while prior models suggest orders of magnitude more traffic~\cite{hong2009analytical,lai2013throughput,lai2013performance,zhou2017performance}. 

Our novel approach separately models the traffic at each level of the memory hierarchy using the access patterns exhibited by the cuDNN library~\cite{cudnn2016userguide}.
Our model accounts for how the computation is blocked for locality and parallelism and how the hardware handles memory accesses in the caches and the software-managed \emph{shared memories}.

We then use the modeled traffic at each memory level and the bandwidths of memories and arithmetic to model overall performance and identify which resource bounds performance under different system configurations. We validate \del with four popular CNNs (AlexNet~\cite{krizhevsky2012imagenet}, VGG~\cite{simonyan2014very}, GoogLeNet~\cite{szegedy2015going}, and ResNet~\cite{he2016deep}) executed on two NVIDIA Pascal GPUs (TITAN Xp and P100) and a Volta GPU (V100), and show that \del is both accurate and robust across the diverse layers and architectures. We also demonstrate how \del can be used for the design-space exploration of future GPUs and identify interesting tradeoffs for efficient CNN execution by independently scaling different GPU resources.

To summarize, our main contributions are:
\begin{itemize}[noitemsep,topsep=5pt,leftmargin=0.15in]
    \item 
        We introduce \del, a GPU performance model for CNNs. Unlike prior work, \del accurately models traffic across all memory hierarchy levels, capturing the data reuse at the different levels; accurately modeling memory traffic is critical for future GPU designs where compute throughput and memory bandwidth must be balanced.
    \item 
        We are first to analyze and model the memory access pattern of the im2col convolution algorithm, which is the most-commonly used algorithm for GPU-accelerated CNNs.
    \item
        We validate the high accuracy and robustness of \del across four popular CNNs and three different GPUs.
    \item 
        We demonstrate how \del can be used to efficiently explore future GPU designs and pinpoint specific resource-scaling opportunities for future GPUs that effectively remove bottlenecks that \del identifies.
\end{itemize}


\section{Background}
\label{sec:background}

\subsection{GPU Architecture}
\label{sub_sec:gpu}
GPUs are designed to accelerate compute-intensive highly-parallel workloads and therefore: (1) require applications to express parallelism with many threads, and (2) contain a large number of very wide SIMD cores, called streaming multiprocessors (SMs). Each SM processes a warp of 32 threads in lockstep, such that ideally all 32 threads execute the same instructions. In addition to the processing elements, each SM contains load store units (LSU), register files (RF), a shared memory RAM (SMEM), and an L1 cache. The SMs share access to the L2 cache and DRAM through a crossbar interconnection network. 

GPU workloads are tiled into thread groups called cooperative thread arrays (CTAs). The CTA scheduling mechanism is assumed to assign SMs to CTAs in a round-robin manner~\cite{lee2014improving}. Each CTA typically consists of multiple thread warps that execute concurrently to hide memory access latencies. Given sufficient resources (RF and SMEM), multiple CTAs can be simultaneously executed within one SM (active CTAs). Interleaving multiple CTAs improves the ability of the SM to perform computation while some warps wait for memory.

\subsection{Convolution Layer Workload}
\label{sub_sec:conv_layer}

CNNs consist of multiple layer types: convolution (\conv) layers extract features from input images, pooling layers reduce feature dimensions, and fully-connected (FC) layers generate the score for each prediction class. Especially, \conv layers have been the major focus of architecture research due to their high computation demands and data reuse~\cite{chen2017eyeriss,kim2016neurocube}. 

\begin{figure}[t!]
    \centering
    \includegraphics[width=0.45\textwidth]{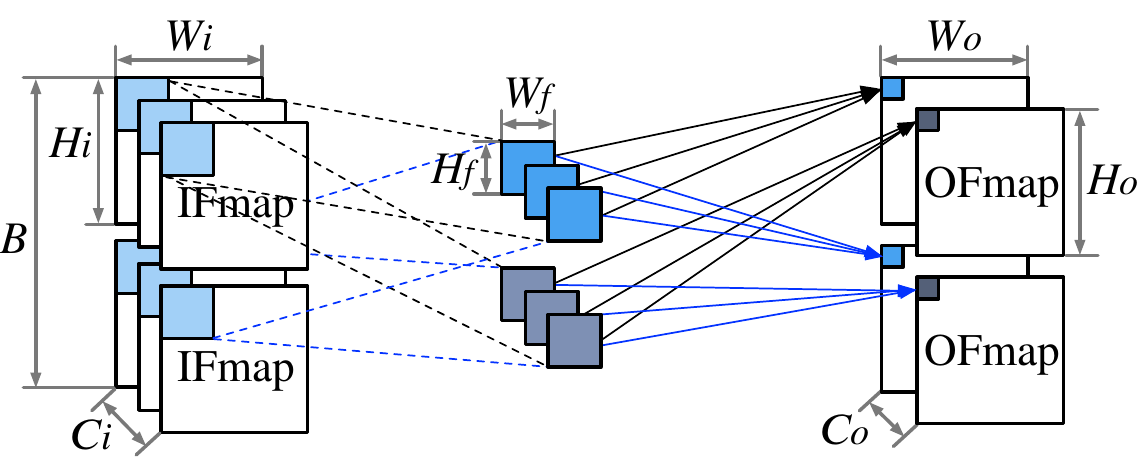}
    \vspace*{-2mm}
    \caption{\small{The dimensions and computation patterns of a \conv layer.}}
    \label{fig:conv_layer}
    \vspace*{-2mm}
\end{figure}

\fig{fig:conv_layer} shows the computation pattern within a \conv layer. To compute a single data element of an output feature map (OFmap), a vector of input feature maps (IFmaps) are convolved by filters (matrices of weights) and accumulated then activated using a non-linear function such as ReLU~\cite{lecun2015deep}. This \conv layer computation is formulated as:
\begin{equation}
\footnotesize
\label{eq:conv}
OFmap[k][x][y] \!=\!\!\! \sum_{c=0}^{C_i-1} \sum_{r=0}^{R-1} \sum_{s=0}^{S-1}Fil[k][c][r][s]{\times}IFmap[c][x+r][y+s]
\end{equation}
\(C,\ H,\ W\) indicate the number of channels, the height and width of each feature and filter, and \(i\), \(o\), and \(f\) denotes input feature, output feature, and filter respectively. As each filter is mapped to a combination of IFmaps and OFmaps so there are \(C_i \times C_o\) filters. \((x,y)\) indicates the position of filter in IFmap and it increments by the filter stride. Also, CNN model is generally trained using a \emph{mini-batch} of samples (\(B\) in \fig{fig:conv_layer}) in parallel which typically ranges 32--512~\cite{keskar2016large}, and the samples in a mini-batch share filters. \eq{eq:conv} also shows the high degree of data-reuse within a \conv layer; all IFmaps, filters, and OFmaps are reused during convolution. Given the high math complexity and abundant data reuse, processing a \conv layer is generally bounded by the compute throughput of a processor.

\begin{figure}[t!]
    \centering
    \includegraphics[width=0.38\textwidth]{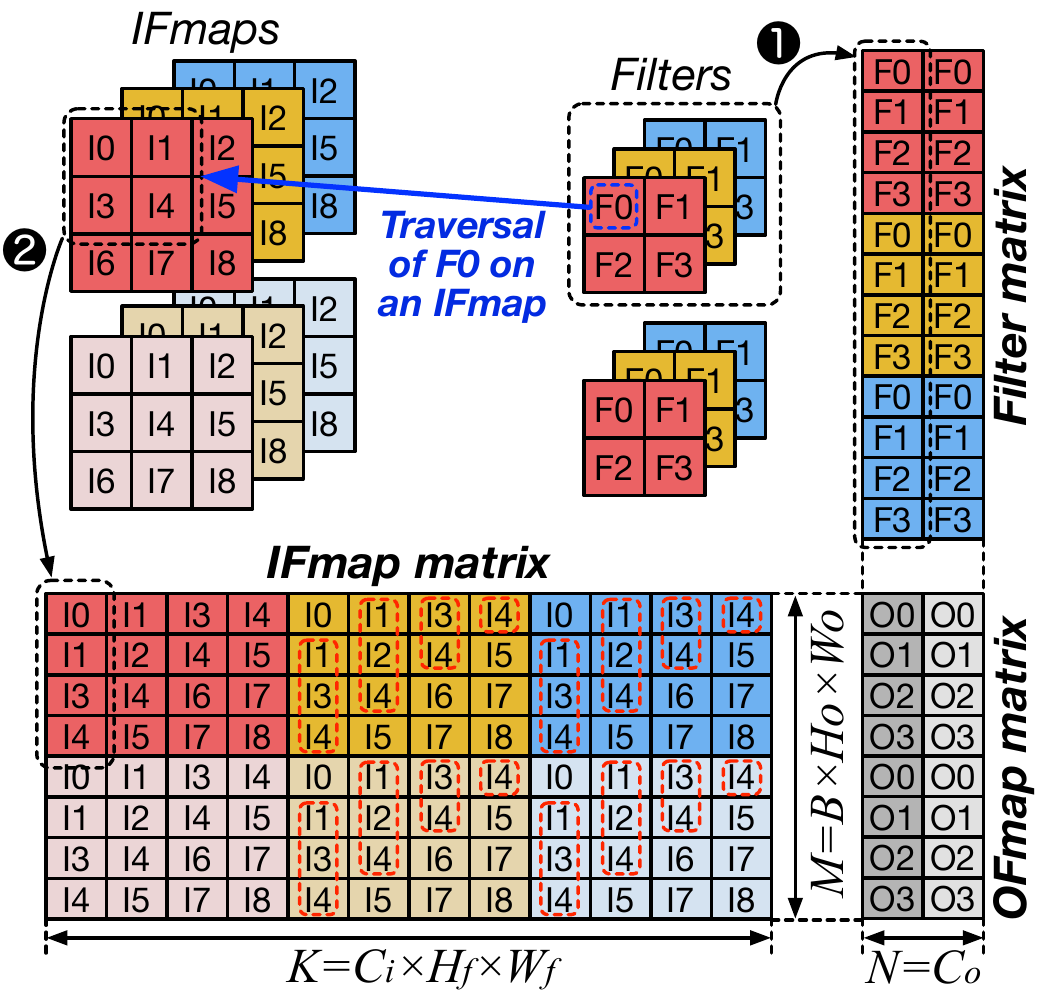}
    \vspace*{-1mm}
    \caption{\small{Im2col GEMM converted from the convolution in \fig{fig:conv_layer}. The red boxed data show duplicated accesses.}}
    \label{fig:im2col}
    \vspace*{-4mm}
\end{figure}

\subsection{Convolution Algorithm \& GEMM Tiling}
\label{sub_sec:conv_lowering}

\textit{\textbf{Parallel Convolution Algorithm.}}
Image-to-column (\emph{im2col}) is one of the most commonly used algorithms for GPU-accelerated convolution kernels~\cite{chetlur2014cudnn,park2018deep} because it works well for a range of \conv layers with different configurations (e.g., different \(B, C, H, W\))~\cite{li2016performance}. To increase data parallelism and GPU resource utilization, im2col transforms the direct convolution described in \fig{fig:conv_layer} into a single general matrix-matrix multiplication (GEMM) with three-dimensions by merging the IFmap and small filter matrices which is illustrated in \fig{fig:im2col}. In this example, first, 2$\times$2 filters for each OFmap channel are converted into columns and stacked \ding{182}. Next, the data elements in the IFmaps are laid out in a way such that the elements to be multiplied by one filter (F0) are placed as a column \ding{183}. The IFmap matrix data layout changes depending on the filter size and the convolution stride. The transformed im2col GEMM has three-dimensions (\(M \times N \times K\)), which are a function of the convolution configuration. Im2col duplicates many data elements which leads to IFmap data reuse in the \conv layer (data in red dotted boxes). Therefore, compared to the typical GEMM, im2col GEMM has greater data reuse and benefits more from caches.

\begin{figure}[t!]
    \centering
    \includegraphics[width=0.48\textwidth]{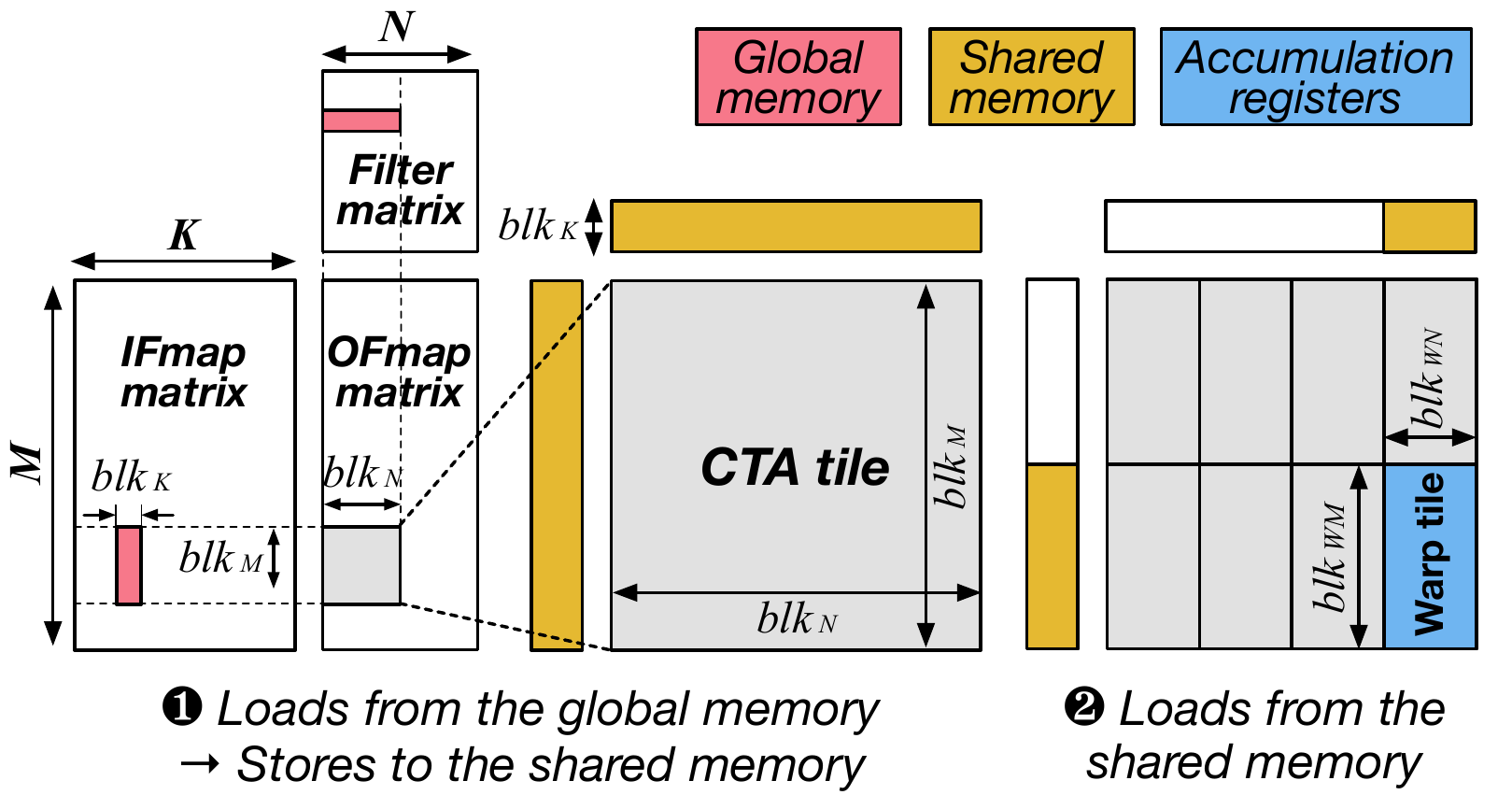}
    \vspace*{-1mm}
    \caption{\small{Im2col GEMM blocking and the data movement for processing a CTA per main loop.}}
    \label{fig:im2col_blk}
    \vspace*{-4mm}
\end{figure}

\textit{\textbf{GEMM Blocking \& Execution Flow.}}
The im2col GEMM is blocked for efficient execution on a GPU (\fig{fig:im2col_blk}). The GEMM blocking divides OFmap matrix with \(M \times N\) dimensions into \(blk_M \times blk_N\) blocks of CTAs (\(blk_M\) and \(blk_N\) are the blocking factors for CTA height and width). Also, GEMM \(K\) dimension is also divided by a blocking factor of \(blk_K\).

Each blocked GEMM execution flow (accumulation in K dimension) consists of three phases: \emph{prologue}, \emph{main loop}, and \emph{epilogue}. 
During the prologue, each CTA loads blocked IFmap \((blk_{M} \!\times\! blk_{K})\) and filter data \((blk_{N} \!\times\! blk_{K})\) from the global memory (DRAM) to registers. GEMM kernels use input double buffering~\cite{kurzak2012autotuning} to overlap these memory loads and the computation routine phases.
Thus, the data loaded in prologue are first fetched to registers and transferred to the shared memory (SMEM) to be used in the first main loop iteration.

The main loops account for the majority of GEMM execution time. The prefetched data in the SMEM in the prior main loop iteration (or prologue) is read and used as the input in the current loop. Specifically, each CTA tile is sub-blocked into warp tiles with a blocking factor of \(blk_{WM} \times blk_{WN}\), and each warp reads both input data from the SMEM. Then, the computation pipeline multiplies the features and weights and accumulates their results in the registers for accumulation. The data loads, computation, and accumulation operations in the main loop routine are software pipelined for efficient resource utilization.

After completing the main loops, at the epilogue stage, the accumulated results are written to the global memory. As all data are written to DRAM, the epilogue can significantly increase the GEMM execution time, particularly when main loops have a small number of iterations.


\section{Motivation and Related Work}
\label{sec:prior_work}

To analyze the performance bottlenecks of GPU-accelerated data parallel workloads, many GPU performance models have been proposed~\cite{hong2009analytical,lai2013throughput,lai2013performance,zhou2017performance,wang2017gpgpu,sim2012performance,wu2015gpgpu}. In particular, prior models do in-depth analysis of the parallel GEMM and show high accuracy~\cite{lai2013performance,zhou2017performance}. These prior models predict application performance based on potential execution time bottlenecks such as computation throughput, instruction fetch/issue slots, and global memory bandwidth. However, they do not model the data traffic at each level of GPU memory system hierarchy which depends on each application's memory access and data reuse patterns. The models proposed by Zhou et al.~\cite{zhou2017performance} and Sunpyo et al.~\cite{hong2009analytical} include cache miss rate as a parameter but it is naively set to 1.
Such assumptions lead to poor performance estimation when the parallel workload has high spatial locality.
CuMAPz{~\cite{kim2011cumapz}} does consider an application's shared memory utilization by analyzing the device code, but does not consider cache and off-chip memory channel traffic.

\begin{figure}[h]
    \centering
    \includegraphics[width=0.46\textwidth]{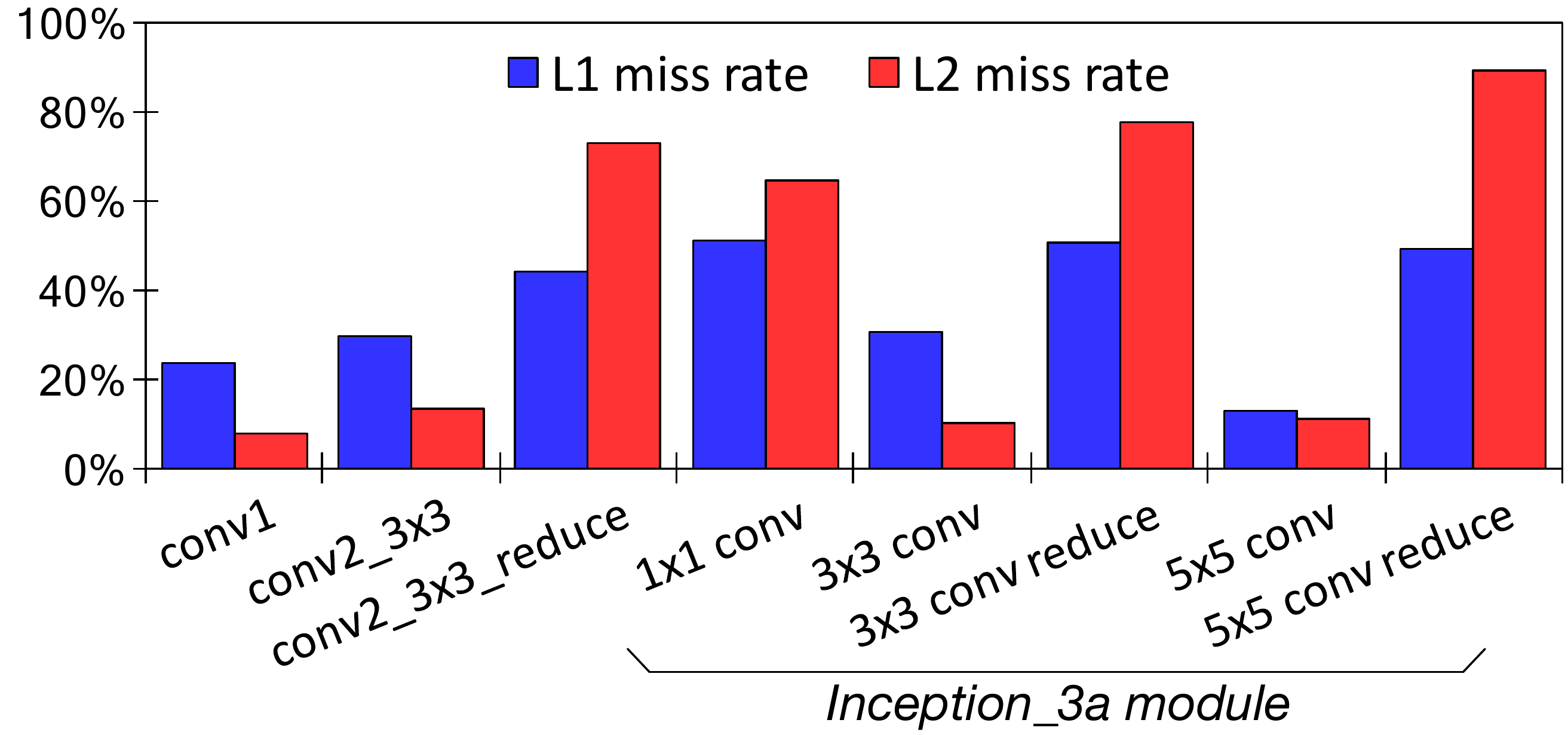}
    \vspace*{-1mm}
    \caption{\small{L1 and L2 cache miss rates of the \conv layers in GoogLeNet}}
    \label{fig:miss_rate}
    \vspace*{-2mm}
\end{figure}

\fig{fig:miss_rate} shows the cache miss rates of \conv layers used in GoogLeNet~\cite{szegedy2015going} measured on NVIDIA TITAN Xp GPU. Depending on the layer configuration, the L1 miss rate varies from 13\% to 50\% and L2 miss rate ranges from 8\% to 90\%. Due to such high traffic variation at different levels of memory hierarchy, the prior performance models fail to accurately predict CNN performance. Particularly, given faster GPU compute throughput scaling than its memory bandwidth, the architecture research for future GPU designs needs accurate data traffic modeling.

\section{Memory Traffic Modeling}
\label{sec:traffic_modeling}

We model the traffic at each level of GPU memory hierarchy using the granularities of data reuse based on the GEMM kernel blocking factors. Our traffic model understands the memory access patterns in the im2col GEMM formed with a specific \conv layer configuration to identify the complex locality within a access granularity at each memory level. We use the implicit convolution kernels supported in cuDNN library, and NVIDIA Pascal GPU as the base architecture. We use 32b floating point data precision widely used for neural network training~\cite{han2015deep}. The minimum memory transaction granularity is 32B, which corresponds to a single sector of one 128B cache line. We use the performance-efficient BCHW as the baseline tensor ordering{~\cite{li2016optimizing,coppens2017gunreal}}.

\subsection{L1 Cache Traffic}
\label{sub_sec:l1_model}

As im2col rearranges the memory access layout, the memory addresses of adjacent IFmap data elements are not continuous. This lowers the spatial locality within each L1 request coalescence granularity (a warp of 32 threads) causing more memory requests than needed. Also, if the memory references are not aligned with the address of the L1 transactions, extra L1 transactions are made. We estimate traffic by calculating this L1 request inefficiency affected by each warp's non-continuous memory references and its address alignment.

\begin{figure}[t]
    \centering
    \includegraphics[width=0.46\textwidth]{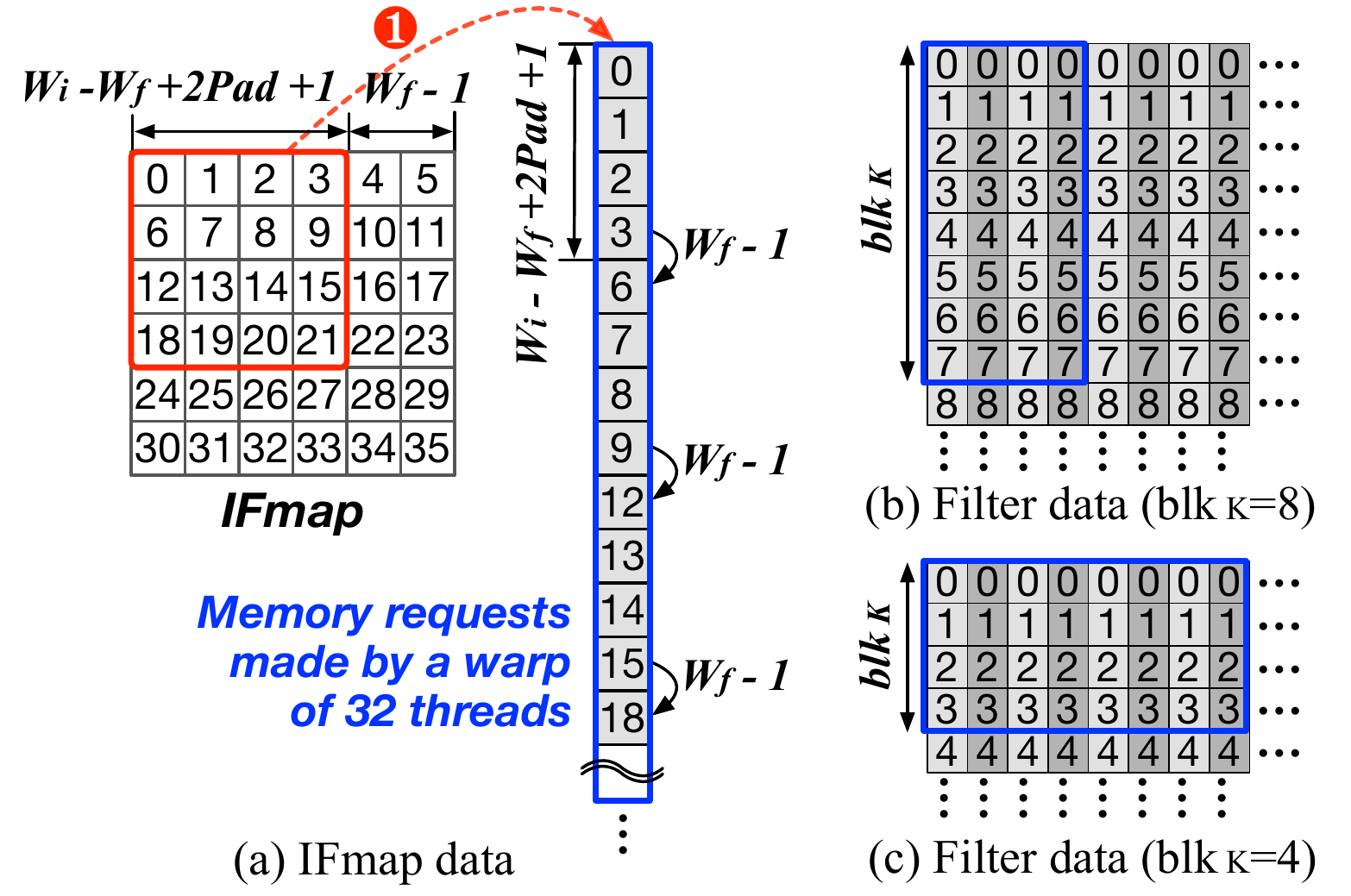}
    \vspace*{-1mm}
    \caption{\small{IFmap and filter data requested by a single warp during a main loop (elements in the blue box). (a) The fraction of IFmap in the red box are the data visited by a single filter data element and it is stacked as a column in the IFmap matrix for GEMM. (b,c) The data layout of the filter matrix for GEMM.}}
    \label{fig:l1_traffic}
    \vspace*{-4mm}
\end{figure}

\fig{fig:l1_traffic}{a} shows the data layout of a single IFmap traversed by one element of a \(3\!\times\!3\) filter with stride 1 and its im2col converted form. Convolution typically adds zero padding (\emph{Pad}) around an IFmap boundary for better feature extraction. In our example, the original IFmap size is \(4 \!\times\! 4\) and it includes pad of 1 which eventually increases the size of IFmap to \(6 \!\times\! 6\). The number on each data element represents the relative location in the physical memory. As described in \sect{sub_sec:conv_lowering}, IFmap data elements visited by a filter data are arranged as a column (\textcolor{red}{\ding{182}}). Then, one warp requests L1 loads for 32 threads which are a fraction of an IFmap column (blue boxed elements in \fig{fig:l1_traffic}{a}). To avoid unnecessary data fetch and save L1 request bandwidth, the L1 loads are coalesced within a warp and this coalescing granularity becomes the fundamental access granularity to L1 cache. This corresponds to the L1 cache line size of 128B (4B $\times$ 32) for Pascal GPU which we experimentally validated with assumption. However, even a warp reads 128B, it cannot be coalesced to a single L1 transaction as data are not continuous: {\small\(W_f\!-\!1\)} elements are skipped every {\small\(W_i\!+2Pad-W_f+1\)} elements (\fig{fig:l1_traffic}{a}). Also, if the stride is larger than 1, data between every two elements are skipped. Reflecting such non-continuous memory access pattern of each IFmap matrix column, its general memory access inefficiency can be calculated as:
\begin{equation}
\scriptsize
\label{eq:access_sparsity}
\begin{split}
\frac{Elements\ requested}{Elements\ used} \!= \! \frac{(W_i+2Pad) \times Strd}{W_i+ 2Pad- W_f+ 1}
\end{split}
\vspace*{-2mm}
\end{equation}
Also, if the coalesced L1 requests made by a warp are not aligned with the L1 transaction addresses, extra transactions are requested. Thus, the overall L1 load inefficiency per warp {\small\((MLI_{IFmap})\)} is calculated by dividing the average L1 requests made per warp by the number of L1 requests per warp with perfect address alignment (\eq{eq:mli_ifmap}).
\begin{equation}
\scriptsize
\label{eq:mli_ifmap}
\begin{split}
MLI_{IFmap} \!&= \!\frac{L1\ requests\ made}{Warp} \times \frac{Warp}{L1\ requests}\\
            \!&= \!\ceil{ \frac{Elmts\ requested}{Elmts\ used} \!\times\! \frac{Bytes}{Warp} \!\!\times\!\! \frac{L1\ requests}{Bytes} } \!\times\! \frac{Warp}{L1\ requests}
\end{split}
\end{equation}

Unlike IFmap, the data elements in the filter matrix are fully continuous in each column. However, based on our profiling result, Pascal GPU uses \(blk_K\) size of 4 or 8 depending on the CTA tile (\fig{fig:l1_traffic}{b}, \fig{fig:l1_traffic}{c}) which results in accessing multiple filter matrix columns by a warp of 32 threads. The data elements in different columns have distant memory addresses as the address is continuous in the column direction. As each 128B L1 request spans beyond the range of filter data at each column, loading an IFmap matrix tile has higher inefficiency \(MLI_{Filter}\) than IFmap's. For Pascal GPUs, considering all possible cache line address alignments, \(MLI_{Filter}\) is calculated as 2.0 and 2.75 when \(blk_K\) is 8 and 4 respectively. Eventually, we calculate the total L1 traffic by multiplying the size of both GEMM inputs (IFmaps and filters) and their MLIs as:
\begin{equation}
\footnotesize
\label{eq:l1_traffic}
\begin{split}
T_{L1\_total}   &= GEMM\_inputs \times MLI\\
                &= (M \! \times \! K) \! \times \! MLI_{IFmap} + (N \! \times \! K) \! \times \! MLI_{Filter}
\end{split}
\end{equation}

\subsection{L2 Cache Traffic}
\label{sub_sec:l2_model}

In im2col GEMM, the IFmap matrix with many duplicated data accesses involves high spatial and temporal locality. As L1 cache is dedicated to a single SM, L1 can capture the reuse within one CTA's IFmap tile but not across active CTAs per SM given its small size ($\sim$32KB). With this assumption, our L2 model estimates the traffic by identifying the unique data accesses made in a CTA input tile.

\begin{figure}[h]
    \centering
    \includegraphics[width=0.48\textwidth]{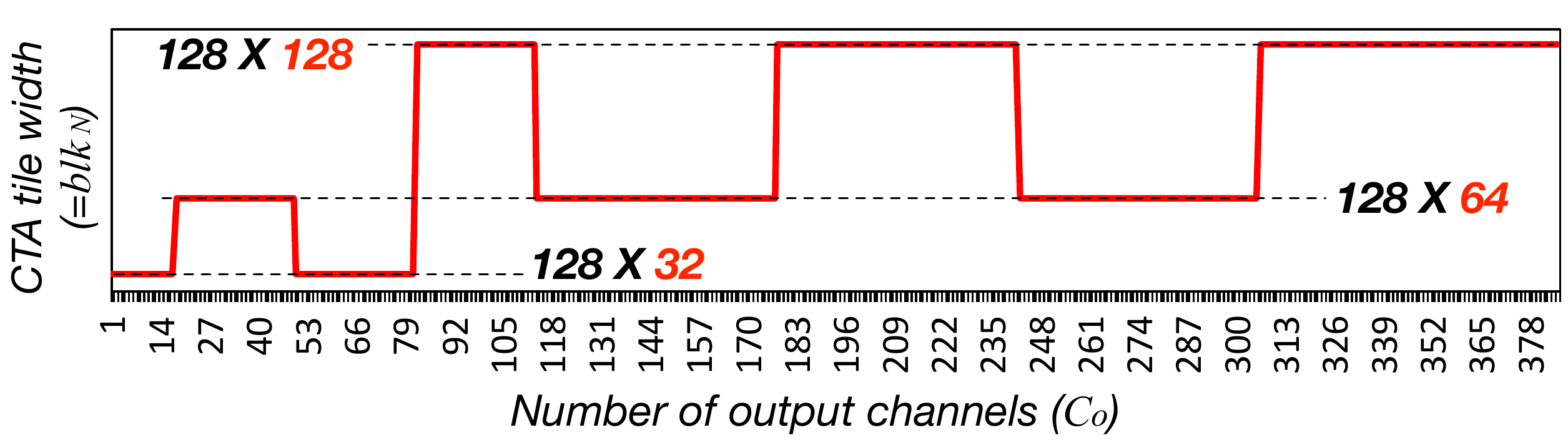}
    \caption{\small{Profiled CTA tile width by different output channel counts.}}
    \label{fig:cta_tile_sizes}
\end{figure}

\textit{\textbf{CTA Tile Selection.}}
CTA tile size affects the spatial locality of the data elements accessed by a CTA; a bigger CTA tile has more data reuse. It also affects the number of CTAs per workload and the ratio between math operations and memory loads per main loop iteration. As it is a critical factor in modeling memory traffic, we analyze the CTA tile size \(((blk_M \times blk_N) \times blk_K)\) of the potential convolution GEMM kernels. Based on our analysis, cuDNN uses three types of CTA tilings: \((128 \times 128)\times 8\), \((128 \times 64)\times 4\), and \((128 \times 32)\times 4\). \(Blk_M\) is fixed to 128 for all tiles and \(blk_N\) and \(blk_K\) are chosen based on the width of the GEMM which equals to \(C_o\). We measure the CTA tile size by different \(C_o\) (\fig{fig:cta_tile_sizes}), and use this as a look-up table to our L2 traffic model.

\begin{figure}[t!]
    \centering
    \includegraphics[width=0.48\textwidth]{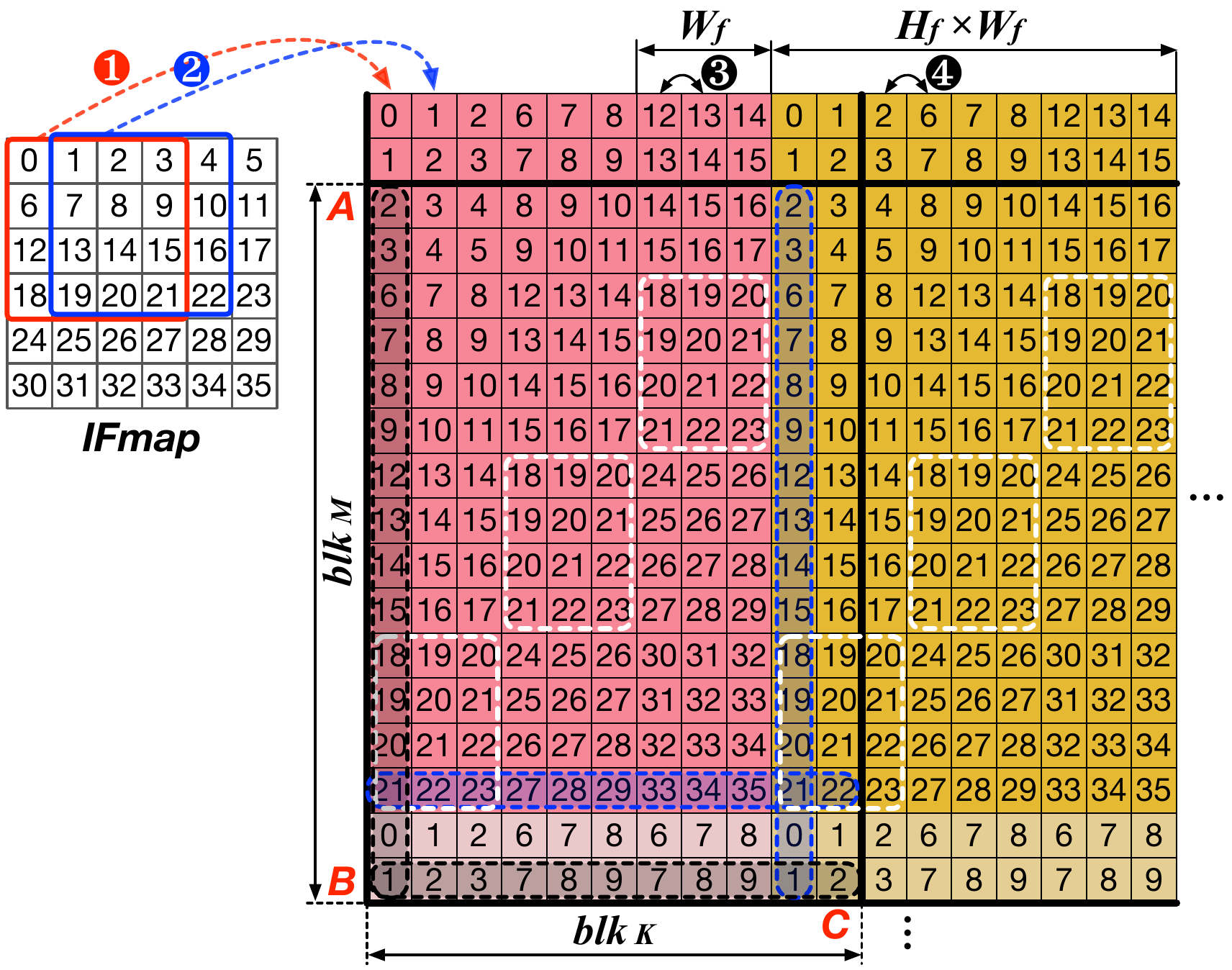}
    \caption{\small{IFmap matrix layout of im2col GEMM. The elements in \(blk_M \!\times\! blk_K \) tile are the input for one main loop. Other details on the figure are described in \sect{sub_sec:l2_model}.}}
    \label{fig:l2_dist}
    \vspace*{-4mm}
\end{figure}

\textit{\textbf{Intra-CTA Spatial Locality.}}
We extend the example used in \sect{sub_sec:l1_model} to explain our L2 traffic model (\fig{fig:l2_dist}). Again, IFmap elements visited by one filter data are remapped as a column in the IFmap matrix (\textcolor{red}{\ding{182}}) and the next column contains the IFmap data traversed by another filter data (\textcolor{blue}{\ding{183}}). Therefore, a \(blk_M \!\times\! blk_K\) IFmap tile for one main loop iteration consists of multiple copies of data that are in continuous address range. This access pattern entails large data locality shown by the duplicated data elements in the white dotted boxes. As L1 cache captures the locality within each tile, only the unique elements are requested to L2. Based on our observation, if we know the address range of data within one IFmap tile (difference between the smallest and the largest address), we can estimate the size of memory requests to L2 in the tile. To be specific, within each IFmap tile, the accessed address increases from the top to the bottom and left to the right. Therefore, using the sum of the address distances between A and B (both edges of \(blk_M\)), and B and C (both edges of \(blk_K\)), we can estimate the total accessed data within an IFmap tile.

We first define the address distance between A and B as {\small\(DIST_V\)} (vertical distance). As the data elements in a column are the IFmap data traversed by a filter data, we formulate {\small\((DIST_V)\)} as \eq{eq:dist_v} using the method used in L1 traffic modeling. Also, when IFmap tile width (\(blk_K\)) spans over more than one channel, multiple unique {\small\(DIST_V\)} can be located within an IFmap tile. In \fig{fig:l2_dist}, \(blk_K\) spans over red and green colored elements each belong to different channel thus an IFmap tile entails two unique {\small\(DIST_V\)} (addition of the blue vertical dotted box). Therefore, using the ratio between \(blk_K\) and a channel width (\(H_f \!\times\! W_f\)), the average vertical distance {\small\((A\_DIST_V)\)} per IFmap can be calculated using \eq{eq:eff_dist_v}.
\begin{equation}
\scriptsize
\label{eq:dist_v}
DIST_{V} = blk_{M} \times \frac{(W_i+2Pad) \times Strd}{W_i+2Pad-Wf+1}
\end{equation}
\begin{equation}
\scriptsize
\label{eq:eff_dist_v}
A\_DIST_{V} = DIST_V \times \frac{blk_K}{H_f \times W_f}
\end{equation}

Next, we define the address distance between B and C of an IFmap tile as {\small\((DIST_H)\)} (horizontal distance). Based on our observation, the distance between two adjacent columns within the same \(W_f\) range is 1 (\ding{184}) as they are the traversals of the two adjacent elements of a filter. However, the distance between the two columns in different \(W_f\) ranges is {\small\(W_i\!+2Pad-W_f+1\)} (\ding{185}) which is the width of the filter element traversal on a IFmap. Also, the alignment of \(blk_K\) to the filter width (\(W_f\)) affects the number of \(W_f\) edges within an IFmap tile; more \(W_f\) edges increases \(DIST_H\). Therefore, considering both the inter-column distances and the alignment of \(blk_K\) to the IFmap layout, we calculate {\small\(DIST_H\)} using \eq{eq:dist_h}. Also, if the elements of more than one data sample are located within a IFmap tile, multiple unique {\small\({DIST_H}\)} can be found within an IFmap tile. In \fig{fig:l2_dist}, the light colored elements belong to a different data sample which adds another unique {\small\(DIST_H\)} (addition of the blue horizontal dotted box). The number of samples per IFmap tile is affected by the ratio between \(blk_M\) and the size of one IFmap which is the product of the height and width of an IFmap or \((\frac{W_i+2Pad-W_f+1}{Strd})^2\) (assuming height and width are the same). Reflecting this, we calculate the average horizontal distance of an IFmap tile {\small\((A\_DIST_H)\)} (\eq{eq:eff_dist_h}).
\begin{equation}
\scriptsize
\label{eq:dist_h}
\begin{split}
    DIST_H  &= \Big( \frac{blk_{K} \!-1}{W_f} \Big) \!\times\! \big( (W_i \!-\! W_f+1) + Strd \!\times\! (W_f \!-\! blk_{K} \!+\! 1) \big)\\
            &+ \Big( \frac{W_f-blk_K+1}{W_f} \Big) \times \big( Strd \!\times\! (blk_K-1) \big)
\end{split}
\vspace*{-3mm}
\end{equation}
\begin{equation}
\scriptsize
\label{eq:eff_dist_h}
A\_DIST_{H} = DIST_H \!\times\! \Bigg( 1+ \frac{blk_M}{\big(\frac{H_i+2Pad-H_f+1}{Strd}\big)^2} \Bigg)
\end{equation}

In case of \(1 \!\times\! 1\) convolution and FC layers, all IFmap data elements are unique and there is no data reuse within an IFmap tile. Therefore, we can simply calculate {\small\(DIST_V\)} as the height of an IFmap tile and {\small\(DIST_H\)} as its width. The same method is applied to filter data loads because they are all unique as well. Finally, using the average distance of IFmap and filter tiles, the total L2 load traffic within a main loop is calculated by adding {\small\(A\_DIST_V\)} and {\small\(A\_DIST_H\)}. Then, the total L2 traffic for a \conv layer is calculated by multiplying the number of CTAs (\(Num_{CTA}\)) and the number of main loop iterations (\eq{eq:l2_traffic}).
\begin{equation}
\scriptsize
\label{eq:l2_traffic}
\begin{split}
T_{L2\_total}   &= A\_DIST_{GEMM\_tile} \times \frac{GEMM\ tiles}{conv\ layer}\\
                &= \big(A\_DIST_{IFmap} + DIST_{Filter} \big) \!\times\! \frac{K}{blk_K} \!\times\! \frac{Num_{CTA}}{conv\ layer}
\end{split}
\end{equation}


\subsection{DRAM Traffic}
\label{sub_sec:dram_model}

\begin{figure}[b!]
    \centering
    \includegraphics[width=0.44\textwidth]{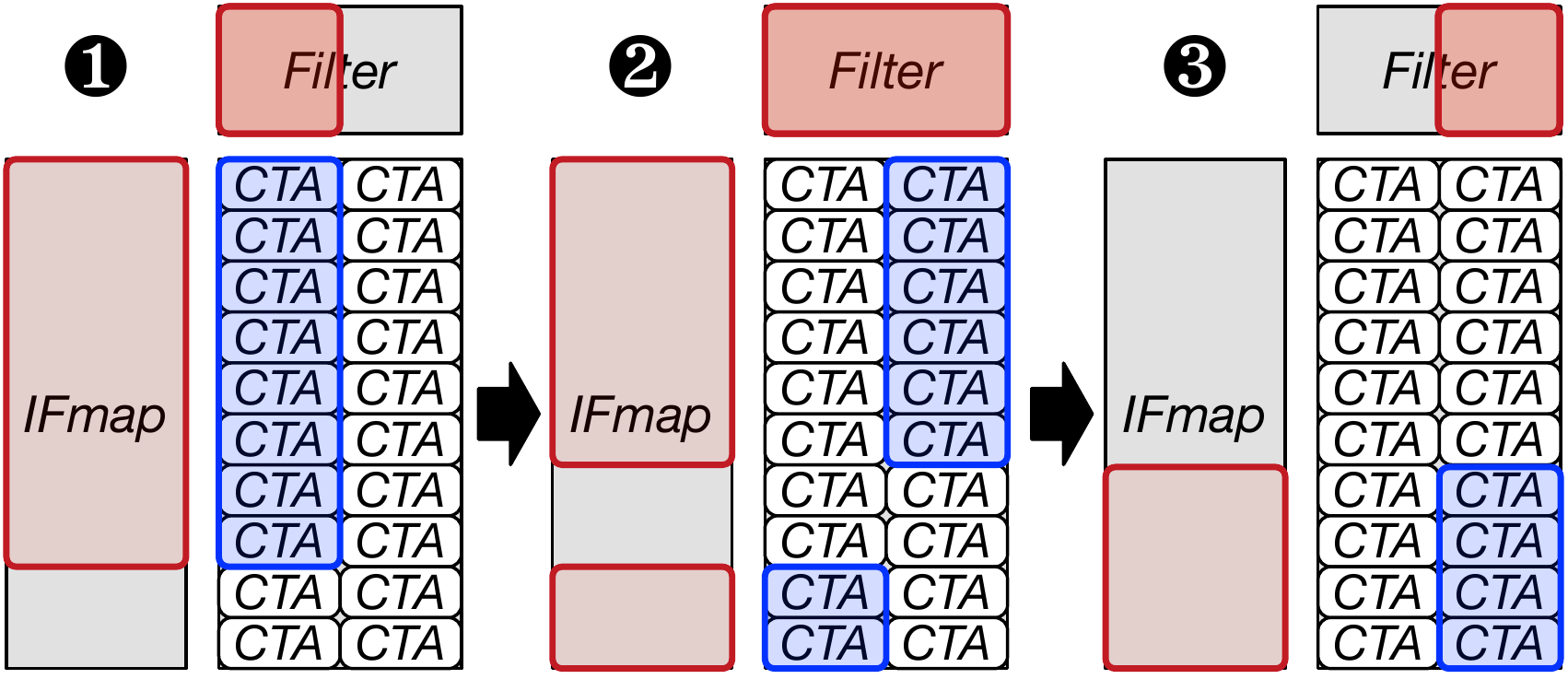}
    \caption{\small{IFmap and filter data reference at sequences of processing CTA batches (\ding{182}, \ding{183}, and \ding{184})}}
    \label{fig:cta_schedule}
    \vspace*{-3mm}
\end{figure}


CTAs in a GEMM kernel share the data in both input matrices. As L2 cache is shared by all SMs, the CTAs executed in parallel by all SMs (a CTA batch) can reuse such shared data. However, the specific inter-CTA data reuse depends on CTA scheduling mechanism. Our DRAM modeling assumes GPU uses column-wise CTA scheduling considering im2col GEMM's skinny shape. Then, we estimate the DRAM traffic by identifying the unique data elements within a CTA batch.

CTA tile array of im2col GEMM is tall as its height is set by the product of the height and width of a feature, and the size of the mini-batch. In contrast, its width is relatively narrow as it equals to the number of output channels. Thus, im2col GEMM CTA tile array has a high aspect ratio making it has many orders of magnitude more CTAs in the column direction than in the row direction. This increases the chance of the CTAs in the same column to be executed in parallel. \ding{182} to \ding{184} in \fig{fig:cta_schedule} illustrate the sequences of processing CTA batches. In this example, the GPU has 20 CTAs and there are 8 SMs so the CTA batch size is 8. At each step, the SMs fetch the red boxed IFmap and filter data.

At the sequences of \ding{182}, \ding{183}, and \ding{184}, many CTAs refer to the same filter data, which makes the filter data have a short reuse distance thus increasing their chance to be resident in L2 cache. Also, each conv layer's total filter size is generally only a few mega-bytes maximum for recent CNNs~\cite{szegedy2015going,he2016deep}, so we effectively consider filter data as loaded from DRAM just once. On the other hand, the IFmaps have long rereference distances between columns of CTA tiles in the CTA tile array. Therefore, the overlapping IFmap data across \ding{182}, \ding{183}, and \ding{184} are fetched twice. This makes the effective IFmap data load count the same as the columns of CTA tiles in the GEMM.
Thus, we calculate the DRAM traffic of IFmap, Filter, and their sum as: 
\begin{equation}
\scriptsize
\label{eq:dram_traffic_boundary_penalty}
\begin{split}
T_{DRAM\_IFmap}     &= IFmap\_size \times \frac{Columns\_of\_CTA\_tiles}{CTA\_tiles\_array}\\
                    &= (B \!\times\! H_i \!\times\! W_i \!\times\! C_i) \times \frac{Columns\_of\_CTA\_tiles}{CTA\_tile\_array}\\
T_{DRAM\_Filter}    &= Filter\_size \\
                    &= (C_i \!\times\! H_f \!\times\! W_f) \times C_o \\
T_{DRAM\_total}     &= T_{DRAM\_IFmap} + T_{DRAM\_Filter}
\end{split}
\end{equation}
Both IFmap height (\(H_i\)) and width (\(W_i\)) are zero padded.
Some IFMap data is no used by a  \(1 \!\times\! 1\) \conv layer with a stride larger than 1 and is not loaded. These data elements are excluded from DRAM traffic.


\section{Performance Modeling}
\label{sec:perf_model}

\del predicts \conv layer's execution time and its performance bottleneck using the memory traffic estimates extracted from our memory traffic model. To identify the execution bottleneck, our model analyzes the compute and memory access streams in the highly software-pipelined GEMM kernel each of which uses different GPU resources.

\begin{figure}[t!]
    \centering
    \includegraphics[width=0.47\textwidth]{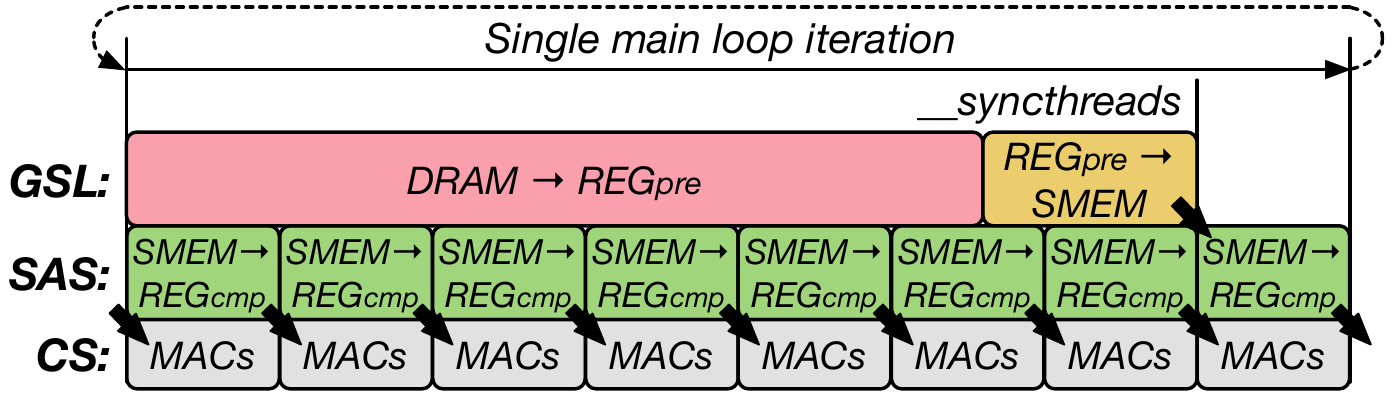}
    \caption{\small{Execution time breakdown of a software pipelined GEMM main loop~\cite{cutlass}. Three execution streams: global load stream (GLS), shared load stream (SAS), and compute stream (CS)}}
    \label{fig:loop_time}
    \vspace*{-3mm}
\end{figure}

\fig{fig:loop_time} shows the execution time breakdown of the software pipelined GEMM main loop~\cite{cutlass} (arrows indicate dependencies between execution blocks). Three execution streams proceed in parallel to maximize resource utilization: the global load stream \emph{(GLS)}, shared memory access stream \emph{(SAS)}, and compute stream \emph{(CS)}, that each exercise a different resource.

First, the global load stream \emph{(GLS)} loads inputs from the global memory to the registers for prefetch and then to SMEM. GLS execution time is determined by both the latency to load data from the global memory and to store it to SMEM. The total load time consists of the (empty) pipeline latency and the transfer latency. The pipeline latency is the data flight time and it includes cache tagging, buffer pipelining, and all circuit data paths. Pipeline latency is fixed regardless of memory traffic, however, data transfer time increases with the data volume because of the limited bus bandwidth. \del calculates GLS execution time by comparing the load latency from L1, L2, and DRAM using the traffic estimated by our memory traffic model (\eq{eq:gls_time}). In \eq{eq:gls_time}, the load latency from each level is the sum of its pipeline latency (\(LAT_{level}\)) and the per-main loop volume (\(TpL_{level}\)) divided by bandwidth (\(BW_{level}\)). \(Num_{SM}\) is the number of SMs per GPU.
\begin{equation}
\scriptsize
\begin{aligned}
\label{eq:gls_time}
t_{GLS}  =  & \max\Big(   LAT_{L1} + \frac{TpL_{L1}}{BW_{L1}},\; LAT_{L2} + \frac{TpL_{L2}}{BW_{L2}/Num_{SM}}, \\
            & LAT_{DRAM} + \frac{TpL_{DRAM}}{BW_{DRAM}/Num_{SM}}\Big)
\end{aligned}
\end{equation}


Second, the shared memory access stream \emph{(SAS)} loads the prefetched data from the SMEM to registers. The execution time of SAS is bound by the SMEM bandwidth and the data volume of both the SMEM loads and the SMEM stores (from GLS). This is because the loads from SMEM share the same data path with the stores to SMEM. The data volume of the SMEM stores is set by the CTA blocking factors \((blk_N+blk_M) \times blk_K\). Then, the SMEM loads transfer the stored data to registers for computation of each warp (Fig 3), thus their data volume is set by the warp blocking factors \((blk_{WN}+blk_{WM}) \times blk_K\) multiplied by the number of warps per CTA (\(Num_{warps}\)). We calculate the SAS execution time per main loop by dividing the data volume of the SMEM stores and SMEM loads by their respective bandwidth \eq{eq:sas_time}.

\begin{equation}
\scriptsize
\label{eq:sas_time}
t_{SAS} =   \frac{(blk_M \!+\! blk_N) \!\times\! blk_K}{BW_{SMEM\_ST}} + 
            \frac{(blk_{WM} \!+\! blk_{WN}) \!\times\! blk_K \!\times\! Num_{warps}}{BW_{SMEM\_LD}}
\end{equation}

\begin{figure}[t!]
    \centering                                                                            
    \subfloat[Simplified CTA GEMM main loop execution timing model]{
        \includegraphics[width=0.48\textwidth]{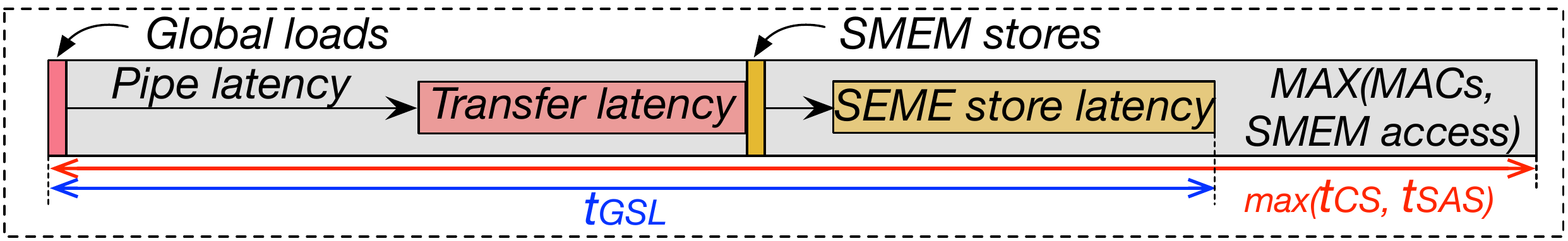}                          
        \label{fig:loop_base}
        
    }\\
    \vspace*{-2mm}
    \centering                                                                            
    \subfloat[Case 1. \(\max(t_{CS},t_{SAS}) \geq t_{GLS}\)]{
        \includegraphics[width=0.48\textwidth]{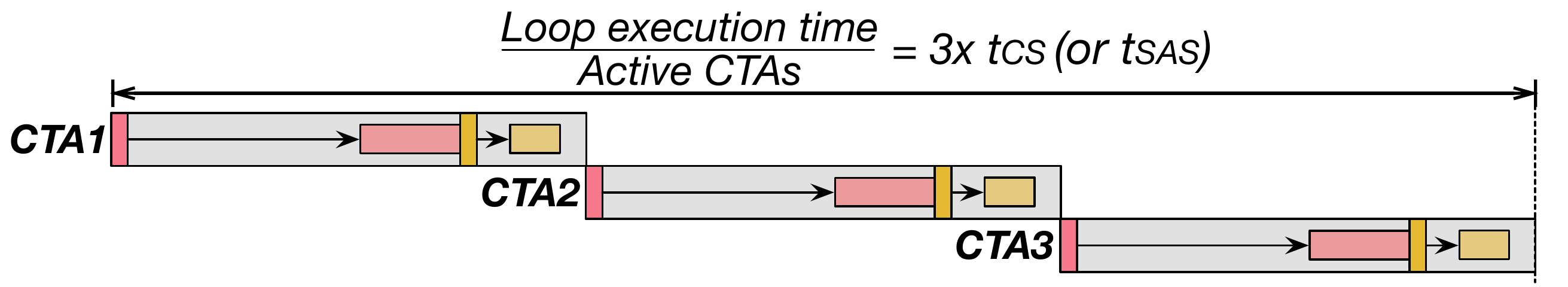}                          
        \label{fig:loop_case1}
    }\\
    \vspace*{-2mm}
    \centering                                                                            
    \subfloat[Case 2. \(t_{GLS} \geq t_{CS} \times Num_{ACT\_CTA}\)]{
        \includegraphics[width=0.48\textwidth]{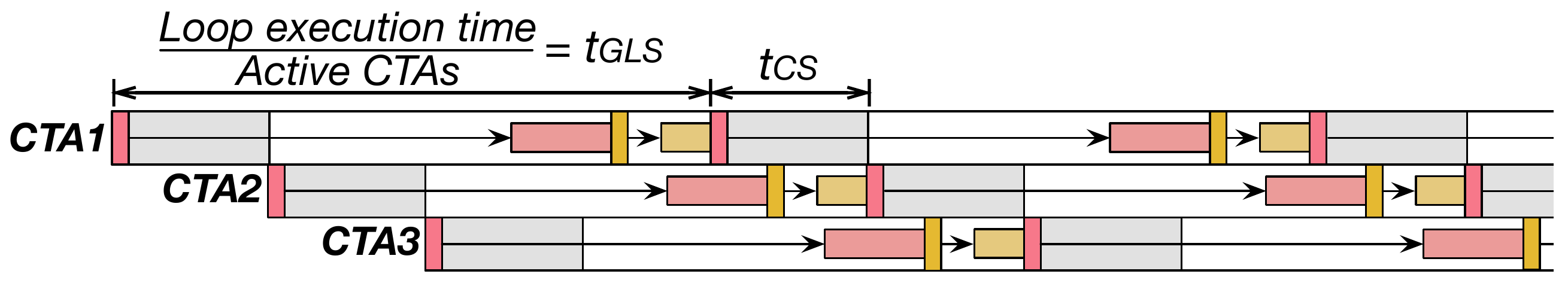}                          
        \label{fig:loop_case2}
    }\\
    \vspace*{-2mm}
    \centering
    \subfloat[Case 3. \(\max(t_{CS},t_{SAS}) \times Num_{ACT\_CTA} \geq t_{GLS} \)]{
        \includegraphics[width=0.48\textwidth]{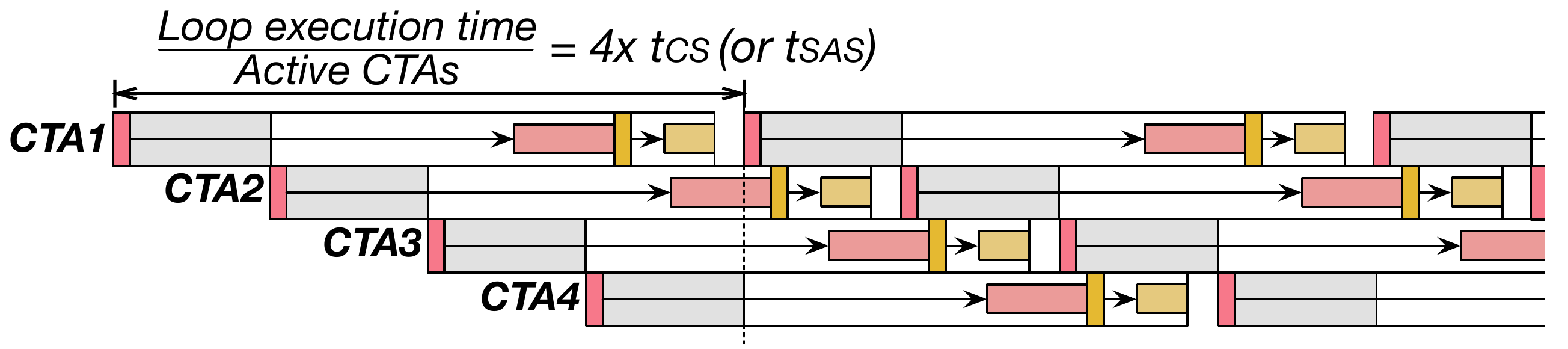}                          
        \label{fig:loop_case3}
    }\\
    \vspace*{-2mm}
    \centering
    \subfloat[Case 4. \(\max(t_{L1\;BW}, t_{L2\;BW}, t_{DRAM\;BW})\) \(\geq t_{CS}\)]{
        \includegraphics[width=0.48\textwidth]{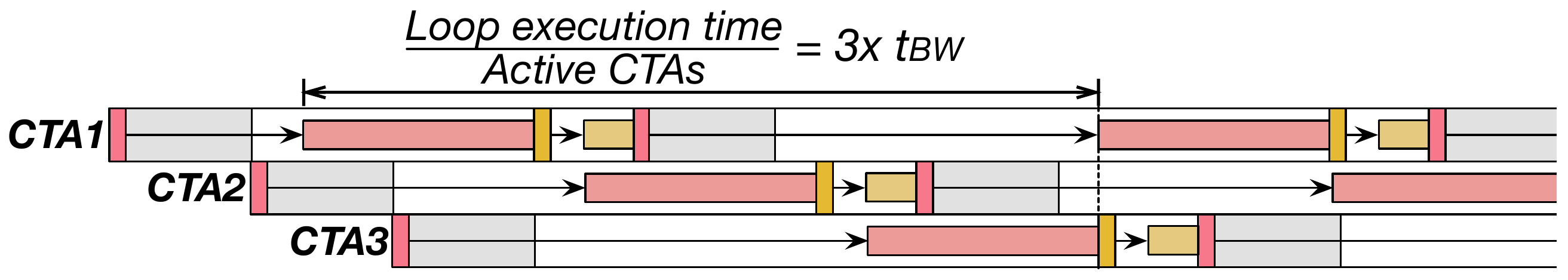}
        \label{fig:loop_case3}
    }\\
\caption{\small{A GEMM loop execution model of active CTAs with different GPU resource bottlenecks.}}
\label{fig:cta_batch_exec_time}
\vspace*{-5mm}
\end{figure}

\begin{table*}[b!]
\renewcommand{\arraystretch}{0.4}
\begin{tabular}{@{}*{2}{p{\dimexpr1\textwidth-\tabcolsep\relax}}@{}}
\footnotesize
\(
t_{Prologue} =  \big( LAT_{DRAM} + \frac{blk_M \times blk_N}{BW_{DRAM}/Num_{SM}} \big) + 
                \big( LAT_{SMEM} +\frac{blk_M \times blk_N}{BW_{SMEM\_ST}} \big) +
                \big( \frac{(blk_{WM} + blk_{WN}) \times blk_K \times Num_{warps}}{BW_{SMEM\_LD}} \big)\) 
                \small\(\;\;\;\;\;(14)\)    
\\ \\
\footnotesize
\(
t_{Epilogue} =  \frac{blk_N \times blk_M}{BW_{DRAM}} ,\;\;
                t_{Epilogue\_bottleneck} = \frac{blk_N \times blk_M}{BW_{Bottleneck(L1,\; L2,\; DRAM)}}\) 
                \small\(\;\;\;\;\;(15)\) 
\\ \\
\footnotesize
\(
t_{MAC_{(sm)}(SMEM_{(sm)})} =   
                t_{Prologue} + 
                \big( t_{CS}(t_{SMEM}) \times \frac{K}{blk_K} + t_{Epilogue} \big) \times \frac{Num_{CTA}}{Num_{SM}}\)
                \small\(\;\;\;\;\;(16)\)
\\ \\
\footnotesize
\(
t_{DRAM\_LAT_{(sm)}} = 
                t_{Prologue} + 
                \big(t_{GLS} + \max\big(\frac{t_{CS}}{blk_K},\frac{t_{SAS}}{blk_K}\big)\big) \times \frac{K}{blk_K} + t_{Epilogue} \big)\times \frac{Num_{CTA}/Num_{SM}}{Num_{ACT\_CTA}}\)
                \small\(\;\;\;\;\;(17)\)
\\ \\
\footnotesize
\(
t_{MEM\_BW_{(sm)}} =    t_{Prologue} + 
                        \big( \max(t_{L1\_BW},t_{L2\_BW},t_{DRAM\_BW}) \times \frac{K}{blk_K} + t_{Epilogue\_bottleneck} \big)
                        \times \frac{Num_{CTA}}{Num_{SM}}\) 
                        \small\(\;\;\;\;\;(18)\)
\vspace*{-20mm}
\end{tabular}
\end{table*}

Finally, the compute stream \emph{(CS)} performs matrix multiplication and accumulation (MAC) operations and its execution time is determined by the compute throughput per SM. The GEMM kernel interleaves pieces of SAS over CS to hide the SMEM access latency. Therefore, if CS is the execution time bottleneck, the loop execution time is the number operations divided by MAC bandwidth (\(BW_{MAC}\)):
\begin{equation}
\small
\begin{split}
\label{eq:cs_time}
t_{CS} = \frac{blk_M \times blk_N \times blk_K}{BW_{MAC}}
\end{split}
\end{equation}

\textit{\textbf{Multi-CTA Interleaving.}}
When a SM has multiple active CTAs, \(t_{GLS}\) can be further hidden by other CTAs' \(t_{SAS}\) or \(t_{CS}\). Especially, the execution time prediction in the context of CTA interleaving is important for GPU with high compute throughput. This is because \(t_{CS}\) per main loop can be shorter than \(t_{GLS}\) thus needs multiple CTAs' \(t_{CS}\) to hide the load latency. The number of active CTAs per SM is set by the ratio between CTA's resource requirement and the size of registers and SMEM per SM. Since the highly-optimized GEMM kernel in cuDNN aggressively reuses registers to increase the number of active CTAs per SM, we use the hardware profiled information for accurate modeling.

Given multiple CTAs to interleave, we model the execution time of the active CTAs with four potential resource bottleneck cases (\fig{fig:cta_batch_exec_time}). The examples (case 1--4) have multiple CTAs to interleave and each row is the time slot for one CTA. We use the simple CTA loop execution time model depicted in \fig{fig:loop_base} as the base. The computes and SMEM accesses are spread over a loop iteration (gray background color).

In case 1, the loop execution time per CTA is bottlenecked by \(t_{CS}\) or \(t_{SAS}\) so we calculate the loop execution time of a CTA batch by adding \(\max(t_{CS},t_{SAS})\) of all active CTAs per SM. Second, if \(t_{CS}\) of all CTAs is shorter than \(t_{GLS}\) but longer than the memory transfer latency, the loop execution time per CTA batch equals to a single CTA's \(t_{GLS}\) (case 2). In this case, each SM has insufficient number of CTAs to hide the loads from the global memory so SM resources are wasted until the entire data arrives for the next loop iteration. Third, if a SM has many CTAs to interleave, \(t_{GLS}\) can be completely hidden by the time for computation or SMEM access (case 3). Finally, if the memory bandwidth becomes the bottleneck due to the high data transfer time (red portion), CTA batch loop execution time is mostly dependent on the data transfer time of the active CTAs (case 4). \del estimates the loop time by comparing the four possible performance bottlenecks.

\textit{\textbf{GEMM Prologue \& Epilogue.}} 
Unlike the GEMM main loop routine, GEMM prologue and epilogue are only constrained by the latency to the global memory or memory bandwidth. For GEMM prologue, considering its small memory access count, we model its impact only for the first CTA assuming the others are hidden by CTA interleaving (Eq. 14). However, GEMM epilogue is not negligible as it has many memory accesses of \(blk_{M} \times blk_{N}\) and all its output are written to the global memory. Therefore, we add the DRAM transfer time of each CTA epilogue to the SM execution time. However, when performance is bottlenecked by the bandwidth of a specific memory level, we use its transfer time (Eq. 15).

Given the execution time of the main loops, prologue, and epilogue, we drive the total execution time of all CTAs allocated per SM by different execution constraints (Eq. 16, 17, and 18). Eq. 16 is used to compute the execution time of case 1 and 3. Eq. 17 and Eq. 18 are used for the case 2 and 4 respectively (Eq. 18 drives three sub-results each for L1, L2, and DRAM BW). Then, the largest value of Eq .16, 17, and 18 becomes the per-SM execution time and its performance bottleneck. Also, as CTAs are not often distributed to all SM, the execution time of the SM with the largest CTAs becomes the eventual \conv layer execution time.

\section{Experiment methodology}
\label{sub_sec:exp_env}

%

\begin{figure*}
    \centering
    \includegraphics[width=1\textwidth]{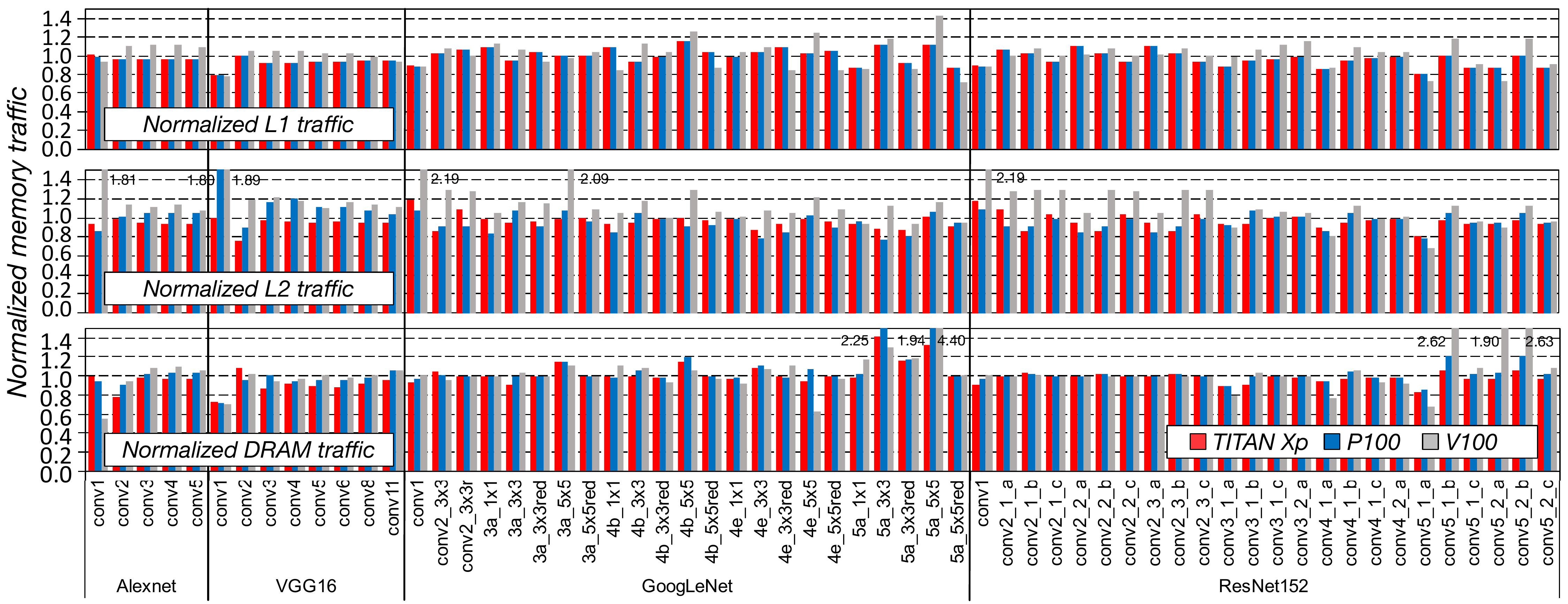}
    \vspace*{-6mm}
    \caption{\small{L1, L2, and DRAM traffic estimates for the \conv layers by \del normalized to the measured values on three different GPUs}}
    \label{fig:cnn_traffic}
    \vspace*{-3mm}
\end{figure*}

\noindent\textit{\textbf{Device Specification.}}
We compare the data traffic and performance estimates to the measured data on two Pascal GPUs (TITAN Xp and P100)~\cite{pascal2016whitepaper} and one Volta GPU (V100)~\cite{volta2017whitepaper} (\tab{tab:device_spec}). The profiled memory traffic and execution cycles are the mean of 10 measurements. Also, since the memory access latencies and bandwidths are not specified by NVIDIA, we measure the access latency and bandwidth to L1, L2, and DRAM using microbenchmarks (both new and from prior work~\cite{mei2017dissecting}).

\begin{table}[h] 
    \caption{GPU device specifications}
    \vspace*{-1mm}
    \centering
    \noindent\resizebox{0.97\linewidth}{!}{
        \tabulinesep=0.5mm
        \renewcommand{\arraystretch}{0.7}
        \begin{tabu}{|[1.2pt]l|l|l|l|[1.2pt]}
            \thickhline
            Specifications & Pascal TITAN Xp & Pascal P100 & Volta V100 \tabularnewline
            \midhline
            \(Num_{SM}\) & 30 & 56 & 84 \tabularnewline
            \hline
            \(Core\;clock\) & 1.58 GHz & 1.2GHz & 1.38GHz\tabularnewline
            \hline
            \(BW_{MAC}\ (FP32)\) & 12134 GFLOPS & 8602 GFLOPS & 14837 GFLOPS \tabularnewline  
            \hline
            \(Size_{REG}\) & 256 KB/SM & 256 KB/SM & 256 KB/SM \tabularnewline
            \hline
            \(Size_{SMEM}\) & 96 KB/SM & 64 KB/SM & $\leq$94 KB/SM \tabularnewline
            \hline
            \(BW_{L1}\) & 92 GB/s/SM & 38.1 GB/s/SM & 94.1 GB/s/SM \tabularnewline
            \hline
            \(BW_{L2}\) & 1051 GB/s & 1382 GB/s & 2167 GB/s \tabularnewline
            \hline
            \(BW_{DRAM}\) & 450 GB/s & 550 GB/s & 850 GB/s \tabularnewline
            \hline
            \(Size_{L2}\) & 3MB & 4MB & 6MB \tabularnewline
            \thickhline   
        \end{tabu}
    } 
    \label{tab:device_spec}
    \vspace*{-1mm}
\end{table} 

\noindent\textit{\textbf{Benchmarks.}}
We evaluate \del on the \conv layers of four popular CNNs (AlexNet~\cite{krizhevsky2012imagenet}, VGGNet~\cite{simonyan2014very}, GoogLeNet~\cite{szegedy2015going}, and ResNet~\cite{he2016deep}) used for ImageNet dataset training and prediction~\cite{ILSVRC15}. Because many \conv layers in these CNNs share configurations, we show the results on the unique subset. Unless specified, a mini-batch size of 256 is used for all evaluated layers. We use cuDNN ConvolutionForward API with {\small IMPLICIT\ PRECOMP\ GEMM} algorithm to run \conv layers compiled with CUDA v8.0 and cuDNN v7.0.


\section{Evaluation}
\label{sec:evaluation}

\subsection{Memory Traffic Model}
\label{sec:evaluation_cnn}

\fig{fig:cnn_traffic} shows \del estimates of data traffic for L1, L2, and DRAM normalized to the measurement of three different GPUs for all unique layer configurations of the 4 evaluated CNNs. Both TITAN Xp and P100 use the same kernels and have an L1 request size of 128B so the measured and predicted L1 traffic is the same for both. \del shows high accuracy with a (7.9\% standard deviation). We are unsure of the L1 request size for Volta and experimented with 32B, 64B, and 128B granularity settings for \del. We observed the best match to measurements with 32B L1 requests and \del matches measured GV100 results with a shows geometric mean absolute error (GMAE) of 6.9\% (13.3\% stdev). 

\textbf{The modeled L2 traffic} has a larger variation in error than L1 traffic. L2 traffic is the result of misses in L1 and our model makes the simplifying assumption that there is no overlap in data access to L1 between different CTAs. We hypothesize that the greater error is indeed the result of multiple concurrent CTAs reusing some data in L1. Two observations from the result strongly support our hypothesis. First, \del over-estimates traffic primarily for layers that have larger features, which lead to greater opportunity for reuse between CTAs. Second, we see larger modeling errors for V100 than P100 and Titan XP. The V100 architecture has larger L1 caches (unified with SMEM)~\cite{volta2017whitepaper} that can also contribute to more reuse across CTAs. Overall, \del is still quite accurate with the GMAE and standard deviation (in parentheses) being 4.2\% (7.2\%) for Titan XP, 6.2\% (14.4\%) for P100, and 12.4\% (25.0\%) for V100.

\begin{figure}[b!]
    \centering
    \includegraphics[width=0.47\textwidth]{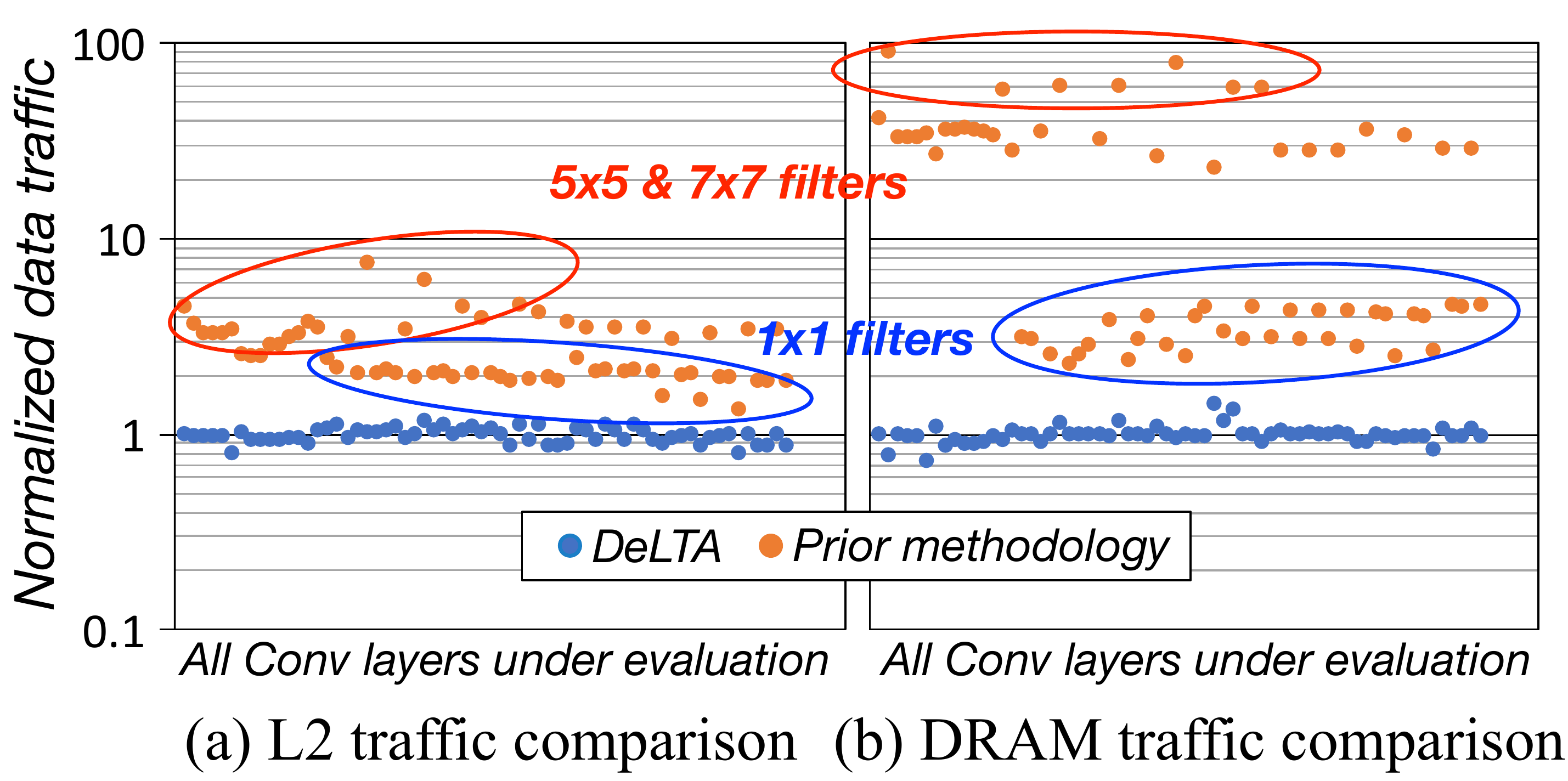}
    \caption{\small{L2 and DRAM traffic estimates by \del and prior methodology normalized to TITAN Xp measurement}}
    \label{fig:traffic_comparison}
    \vspace*{-3mm}
\end{figure}

\begin{figure*}[b!]
    \centering
    \includegraphics[width=0.99\textwidth]{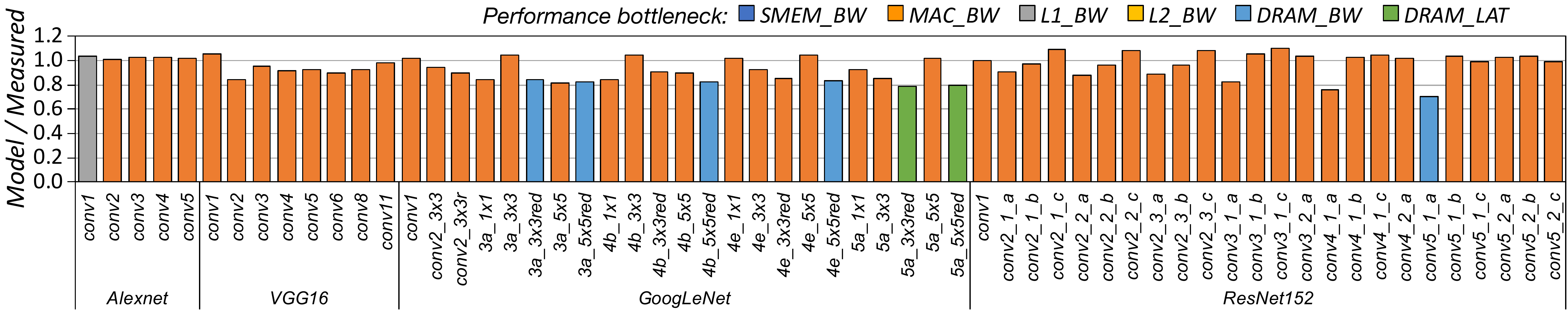}
    \vspace*{-1mm}
    \caption{\small{Conv layer execution time estimates by \del normalized to TITAN Xp's and their performance bottlenecks}}
    \label{fig:performance_titan_xp}
    \vspace*{-3mm}
\end{figure*}

\begin{figure*}[b!]
    \centering
    \includegraphics[width=0.99\textwidth]{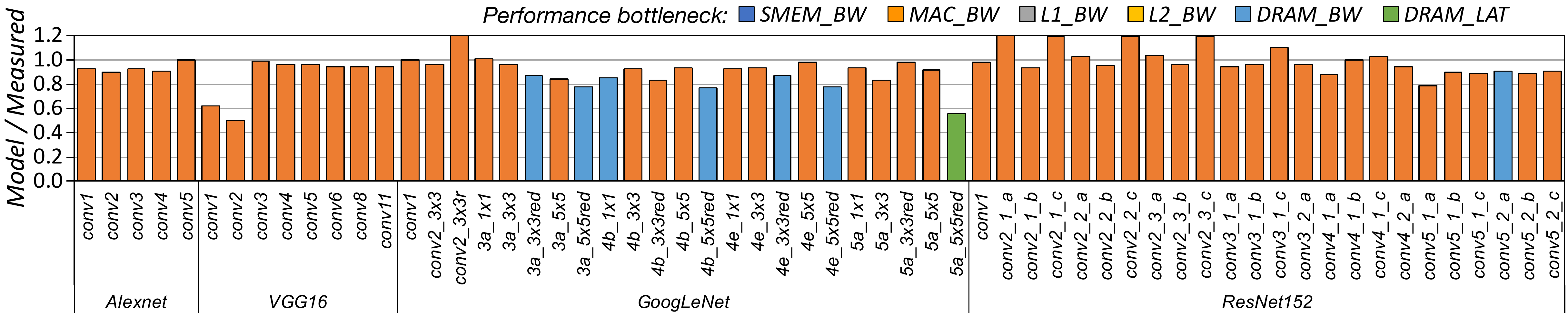}
    \vspace*{-1mm}
    \caption{\small{Conv layer execution time estimates by \del normalized to TESLA V100's and their performance bottlenecks}}
    \label{fig:performance_titan_v100}
    \vspace*{-3mm}
\end{figure*}

\textit{\textbf{The modeled DRAM}} traffic is very accurate overall with a few notable outliers. Some of the layers in GoogLeNet and and ResNet have very small memory footprints that can completely fit within the L2 cache. The profiler we use to measure results reports anomalous numbers for these layers that suggest the impossible scenario where less data is read from DRAM than the actual footprint, leaving \del with a large over-estimation. We attribute this to measurement errors and ignore these $>2\times$ errors from the average numbers reported below. The other source of large modeling error relates to L2 cache behavior and CTA scheduling. \del underestimates DRAM traffic for VGG16-conv1, GoogLeNet-4e\_5{$\times$}5, and a few ResNet152 layers. Our analysis indicates that these errors result from \del identifying potential data reuse in L2 between CTAs that is not exploited by the hardware. For example, VGG16-conv1  uses large IFmaps \(224 \!\times\! 224\) with a small filter stride of 1. This leads to a large overlap between layers given \del's assumptions on scheduling, which do not manifest in practice. Overall, the DRAM traffic model shows small GMAE (with standard deviation) of 2.8\% (10.3\%) for Titan Xp, 6.2\% (14.4\%) for P100, and 10.2\% (9.2\%) for V100 (without the anomalous  measurements).

\textit{\textbf{Memory Traffic Comparison with Prior Models.}}
\fig{fig:traffic_comparison} compares the normalized data traffic for all unique \conv layers under evaluation for \del and the prior models~\cite{zhou2017performance,hong2009analytical}. As the prior models assume 100\% cache miss rates for both levels of caches, we apply the L1 load traffic to both L2 and DRAM. Both L2 and DRAM traffic assumed by the prior models are far from the measurements because these prior models ignore the high data reuse in the \conv layers. The deviation is relatively small for layers with \(1 \!\times\! 1\) filters due to the low data reuse but the large filters have very large errors. These errors are multiple factors (up to nearly 100$\times$) larger than those of \del, and lead to wrong conclusions about performance bottlenecks of modern GPUs with their large arithmetic to memory throughput ratios.

\subsection{Performance Model}
\fig{fig:performance_titan_xp} and \fig{fig:performance_titan_v100} show the performance estimations vs.~ the measured data on TITAN Xp and TESLA V100 respectively, along with their bottlenecks. The GMAE is 6.0\% and 6.5\% for TITAN Xp and V100 respectively. Although highly accurate, \del underestimates the execution time for some layers regardless of their bottlenecks. One major reason is that the estimated memory traffic is uniform across each CTA's GEMM main loop but the actual data traffic is not. For example, if the data fetched by one iteration is reused in the next, the first iteration  execution time is bound by the data loads but the second iteration is constrained by arithmetic throughput. Such non-uniform execution time bottlenecks make \del provide conservative estimates that represent worst-case performance.

Our evaluation shows that arithmetic throughput is the major performance bottleneck (90\% of evaluated layers), which is expected due to the high data reuse of im2col GEMM. We also observe that some layers are bottlenecked by other resources. L1 BW restricts the first \conv layer of AlexNet on TITAN Xp due to its poor L1 transaction efficiency. Many layers in GoogLeNet are bottlenecked by DRAM BW or latency. The layers bottlenecked by DRAM latency do not have enough CTAs to hide the load latency. The layers restricted by DRAM BW have more CTAs to interleave thus saturate the DRAM channels.

\begin{figure}[h]
    \centering
    \includegraphics[width=0.48\textwidth]{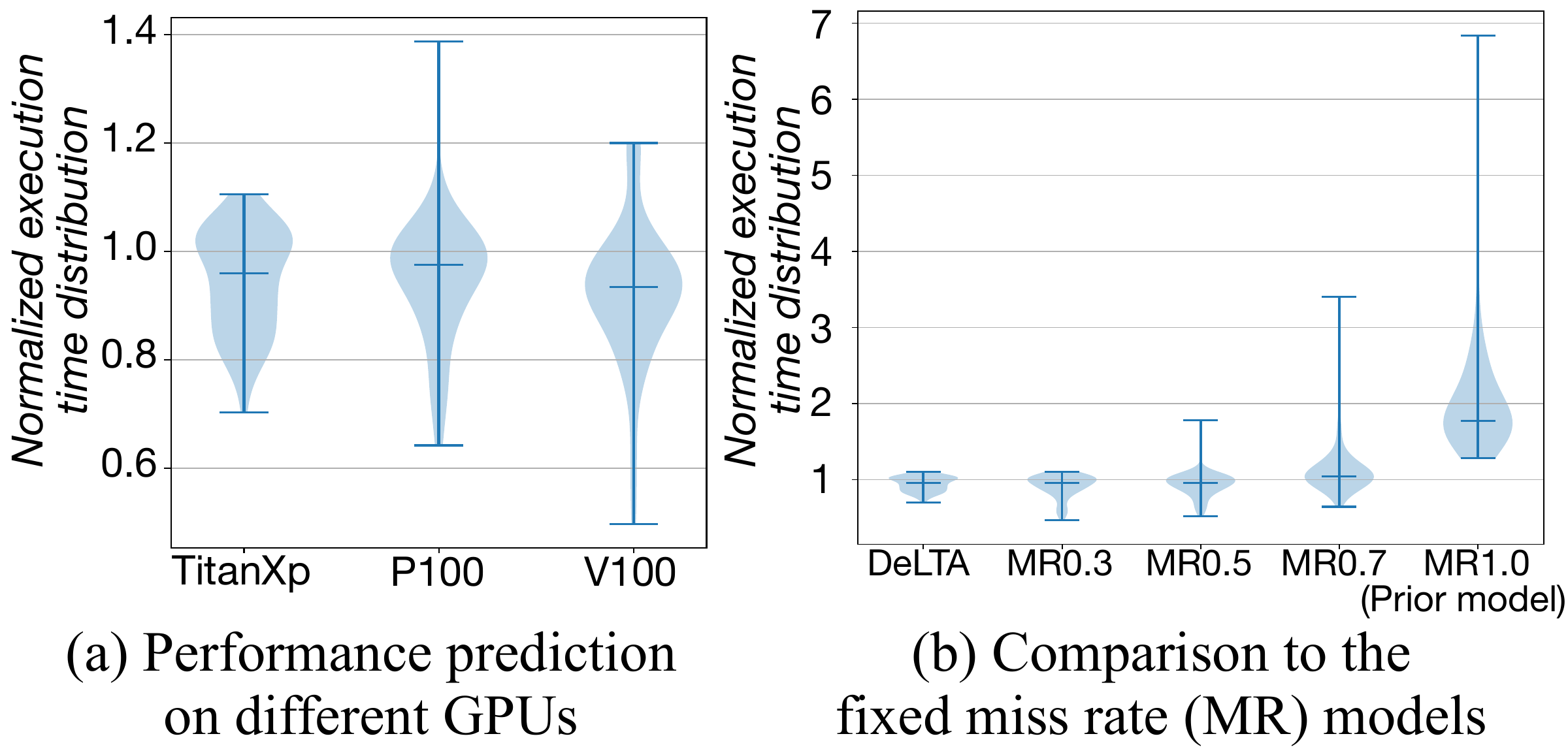}
    \caption{\small{Execution time estimates normalized to TITAN Xp's}}
    \label{fig:performance_comp_all}
    \vspace*{-3mm}
\end{figure}

\textit{\textbf{Performance Estimation with Different GPUs.}}
\fig{fig:performance_comp_all}{a} compares the performance estimation distribution for the three GPUs. \del estimates performance best for Titan XP, but the overall accuracy is quite good For P100 and V100 as well with a robust low-variance estimation (10\% standard deviation). The outliers correspond to those layers for which data traffic is poorly estimated because of dynamic behavior, as explained above. 

\textit{\textbf{Performance Model Comparison.}}
\fig{fig:performance_comp_all}{b} compares the normalized performance estimation results of \del and the models; while prior work advocated using a 1.0 miss rate, we sweep a range of miss rates in the figure. Compared to \del, the other models show wider estimation errors and a larger number of outliers. With the 1.0 miss rate advocated by the prior models, the layer execution time is over-predicted by 1.8$\times$ on average and up to $7\times$. The prediction error for the models using fixed miss rates becomes significantly larger when compute throughput scales as many layers become memory system resource bottleneck.

\subsection{GPU Scaling Study}
We use \del to explore GPU designs for efficient CNN performance scaling with less HW resources. We use the entire 152 \conv layers in ResNet152 to evaluate the potential speedup over the Titan Xp baseline. As almost all compute kernels of ResNet are \conv layers, its performance is dominated by their execution time. \tab{tab:design options} shows the design options used in our experiment. Option 1 and 2 represent the conventional way to improve GPU performance, which keeps SM resources constant and scales the number of SMs and L2 and DRAM BW. These options are expensive as each extra SM involves the entire SM resources such as registers, SMEM capacity and BW, and L1. For the other design options, we use the resource bottleneck information from \del to minimally  scale independent resources for efficient performance gain.

\begin{figure}[t!]
    \centering
    \setlength\tabcolsep{3pt}
    \subfloat[GPU design options. Each column indicates a GPU design choice and xX indicates the magnitude of increase.]{
    \label{tab:design options}
    \noindent\resizebox{0.95\linewidth}{!}{
        \tabulinesep=0.5mm
        \centering
        \renewcommand{\arraystretch}{0.7}
        \begin{tabu}{|[1.2pt]c|[1.2pt]c|c|c|c|c|c|c|c|c|c|[1.2pt]}
             \thickhline 
            Resources & TitanXp & 1 & 2 & 3 & 4 & 5 & 6 & 7 & 8 & 9   \tabularnewline
            \thickhline 
            \(N_{SM}\)     & 1X & 2X & 4X & 1X  & 1X  & 1X  & 1X  & 1X  & 2X & 1X      \tabularnewline
            \hline
            \(MAC_{BW}\) / \(SM\)   & 1X & 1X & 1X & 2X & 4X & 4X & 6X & 8X & 4X & 8X \tabularnewline
            \hline
            \(REGS_{SIZE}\) / \(SM\) & 1X & 1X & 1X & 1X & 1X & 2X & 2X & 3X & 2X & 3X \tabularnewline
            \hline
            \(SMEM_{SIZE}\) / \(SM\) & 1X & 1X & 1X & 1X & 1X & 2X & 2X & 3X & 2X & 3X \tabularnewline
            \hline
            \(SMEM_{BW}\) / \(SM\)  & 1X & 1X & 1X & 1X & 1X & 2X & 2X & 3X & 2X & 3X \tabularnewline
            \hline
            \(L1_{BW}\) / \(SM\)    & 1X & 1X & 1X & 1X & 1X & 1.5X & 2X & 2X & 2X & 2X  \tabularnewline
            \hline
            \(L2_{BW}\)    & 1X & 1.5X & 2X & 1X & 1X & 1.5X & 1.5X & 2X & 2X & 2X  \tabularnewline
            \hline
            \(DRAM_{BW}\)  & 1X & 1.5X & 2X & 1X & 1X & 1.5X & 2X & 2X & 2X & 3X  \tabularnewline
            \hline
            CTA tile H,W   & 128 & 128 & 128 & 128 & 128 & 128 & 128 & 256 & 256 & 256  \tabularnewline
            \thickhline   
        \end{tabu}
        } 
    }\\
    \vspace*{2mm}
    {
        \includegraphics[width=0.48\textwidth]{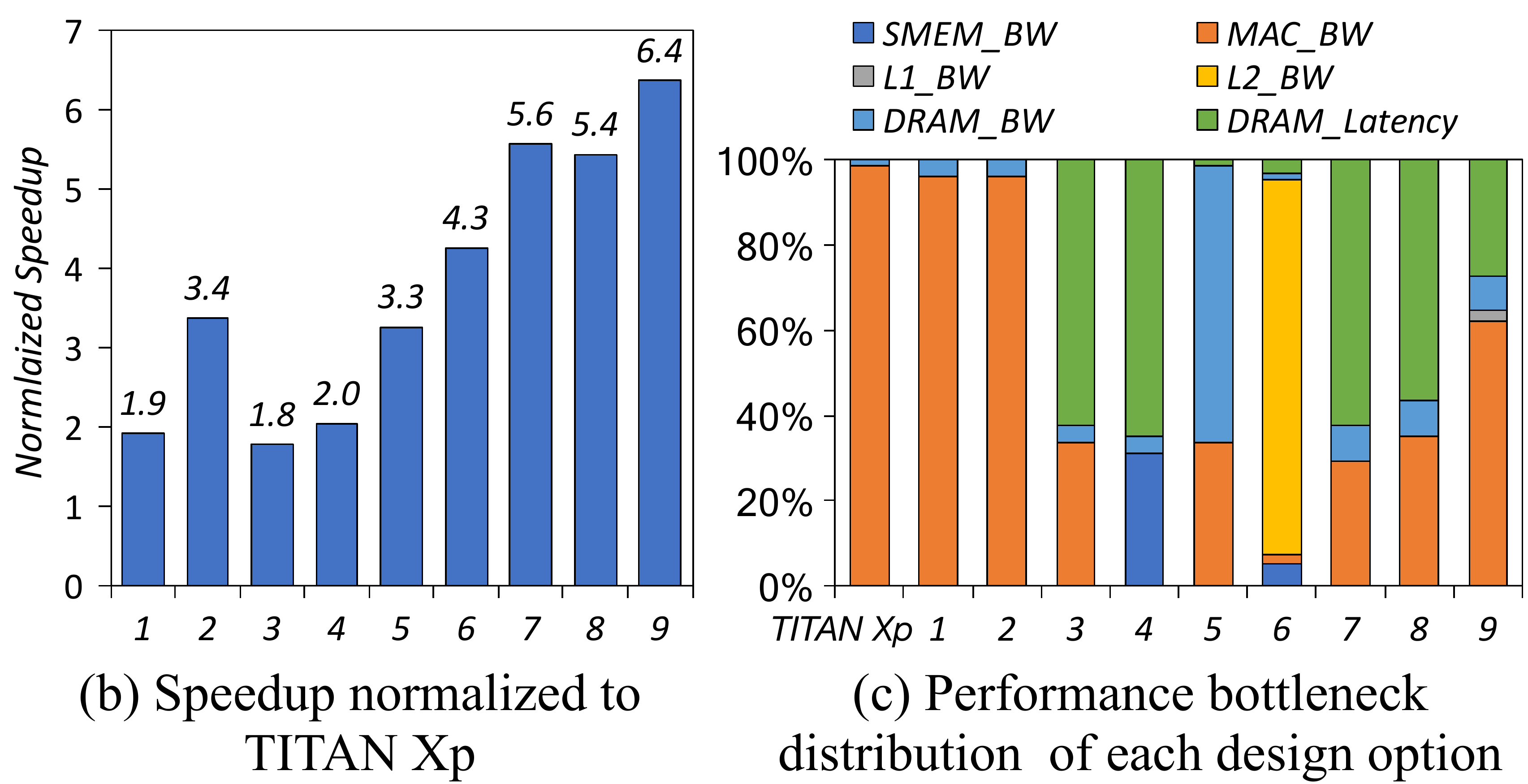}
    }\\
    \caption{\small{GPU resource scaling and speedup of \conv layers in ResNet}}
    \label{fig:resnet}
    \vspace*{-3mm}
\end{figure}

\fig{fig:resnet}{b} shows the normalized performance gain for the different design options  with their performance bottlenecks shown in \fig{fig:resnet}{c}. \del predicts that increasing the number of SMs by 2$\times$ and 4$\times$ along with L2 and DRAM BW improves ResNet forward propagation performance by 1.9$\times$ and 3.4$\times$, respectively. The  limiting factor is  memory BW increase, which restricts pipelined GEMM epilogue and some layers with larger memory BW requirements. Given \conv layers are compute throughput hungry, options 3 and 4 increase only the per core arithmetic throughput by adding more MAC units. However, their performance headroom is only 2$\times$ with most layers bottlenecked by DRAM BW and SM resources.

Based on these observations, option 5 minimally increases resources to avoid bottlenecks, showing similar performance gains to option 2 with far lower hardware resource increases. Option 6 further increases compute throughput but \del shows that now L2 BW becomes the limiter. From option 7 to 9, we increase the size of the GEMM tiles given the high compute throughput to memory bandwidth ratio. Such GEMM parameter choices are only beneficial for GPU designs with high arithmetic throughput, otherwise the performance is restricted by the memory system BW. Design option 8  increases the number of SMs by 2$\times$, but \del shows that increasing DRAM BW is more beneficial than doubling the SM resources (option 9). 

Using \del and a model of hardware resource costs, optimizing a future GPU for CNNs becomes a convex optimization problem. However, modeling precise hardware resource costs is outside the scope of the paper.



\section{Conclusion}
\label{sec:general}

In this paper, we introduce \del which accurately models the memory traffic of a convolution layer at all memory hierarchy levels of a GPU. This is novel and unique in both accounting for the complex memory reuse and execution patterns of im2col which is the most-commonly used convolution algorithms for accelerating CNNs on GPUs. We use the traffic estimates to model the expected performance of a convolution layer on a GPU, where all important GPU characteristics are parameterized to enable the rapid identification of bottlenecks and the evaluation of design tradeoffs. We show how this can be used to explore the design space and better tune resource provisioning for CNNs when compared to equally scaling all resources or ignoring the need for higher memory bandwidth as arithmetic throughput increases.


\section{Acknowledgment}
\label{sec:acknowledgment}

The authors would like to acknowledge NVIDIA for providing GPU resources.


\bibliographystyle{IEEEtran}
\bibliography{main.bbl}
\clearpage
\section*{APPENDIX}
\label{sec:appendix}

\subsection{Memory Traffic Estimates Sensitivity to Convolution Configurations}
\label{subsec:evaluation_sensitivity}

\begin{figure}[h]
    \centering
    \subfloat[Sensitivity to output channel count \((=C_o)\)]{
        \includegraphics[width=0.47\textwidth]{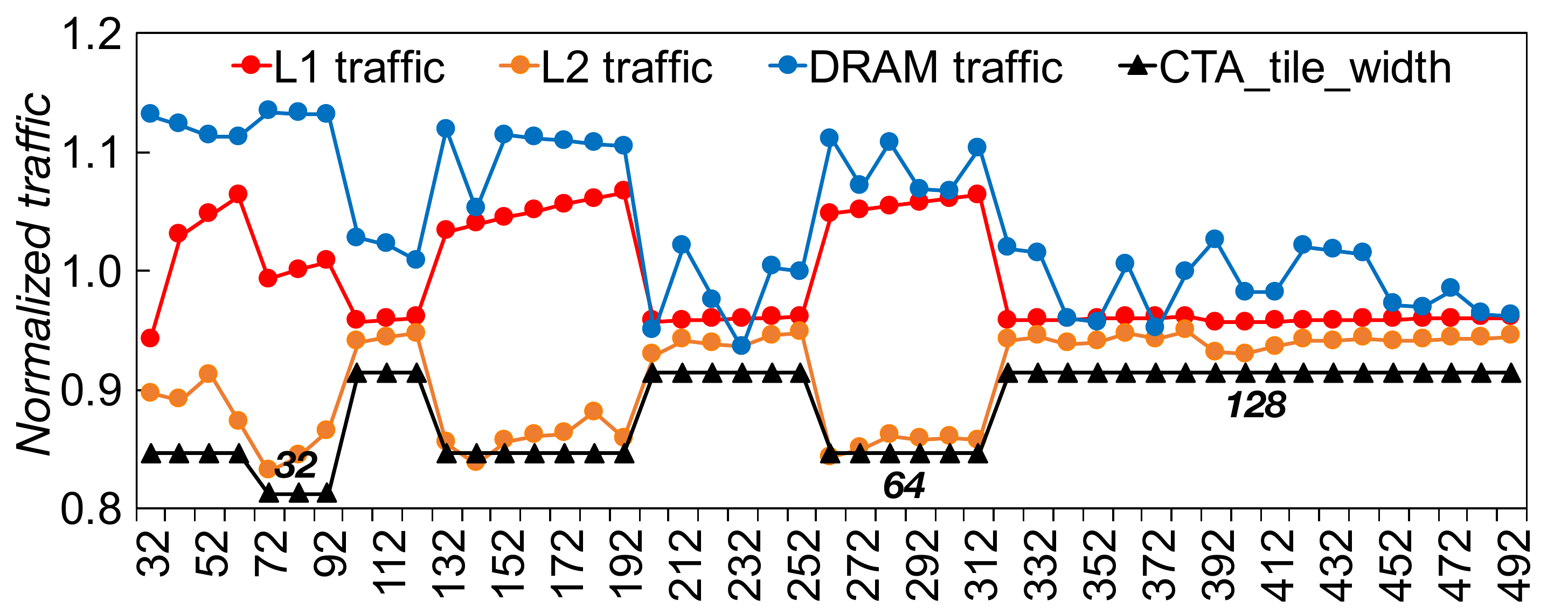}
        \label{fig:sensitivity_K}
    }\\
    \centering                                                                            
    \subfloat[Sensitivity to input channel count \((=C_i)\)]{
        \includegraphics[width=0.47\textwidth]{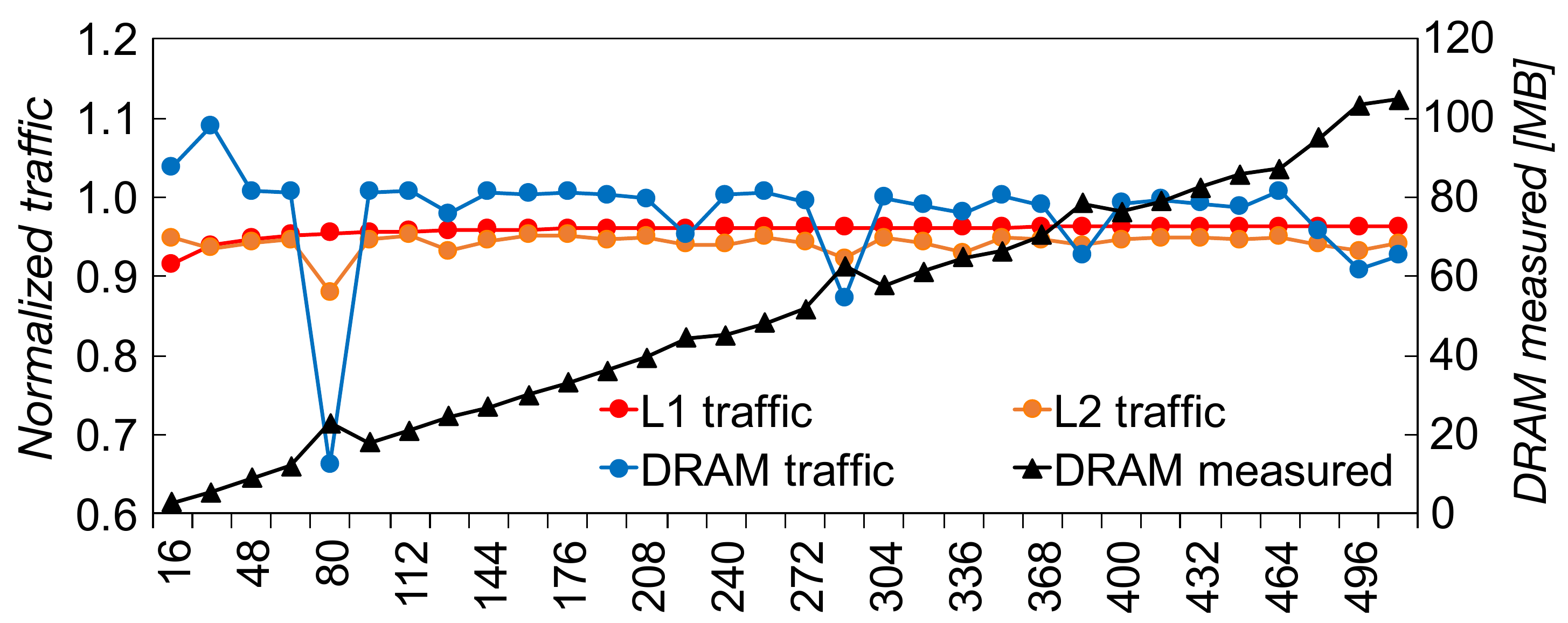}                          
        \label{fig:sensitivity_C}
    }\\
    \vspace*{-1mm}
    \centering
    \subfloat[Sensitivity to IFmap size \((=H_i, W_i)\)]{
        \includegraphics[width=0.47\textwidth]{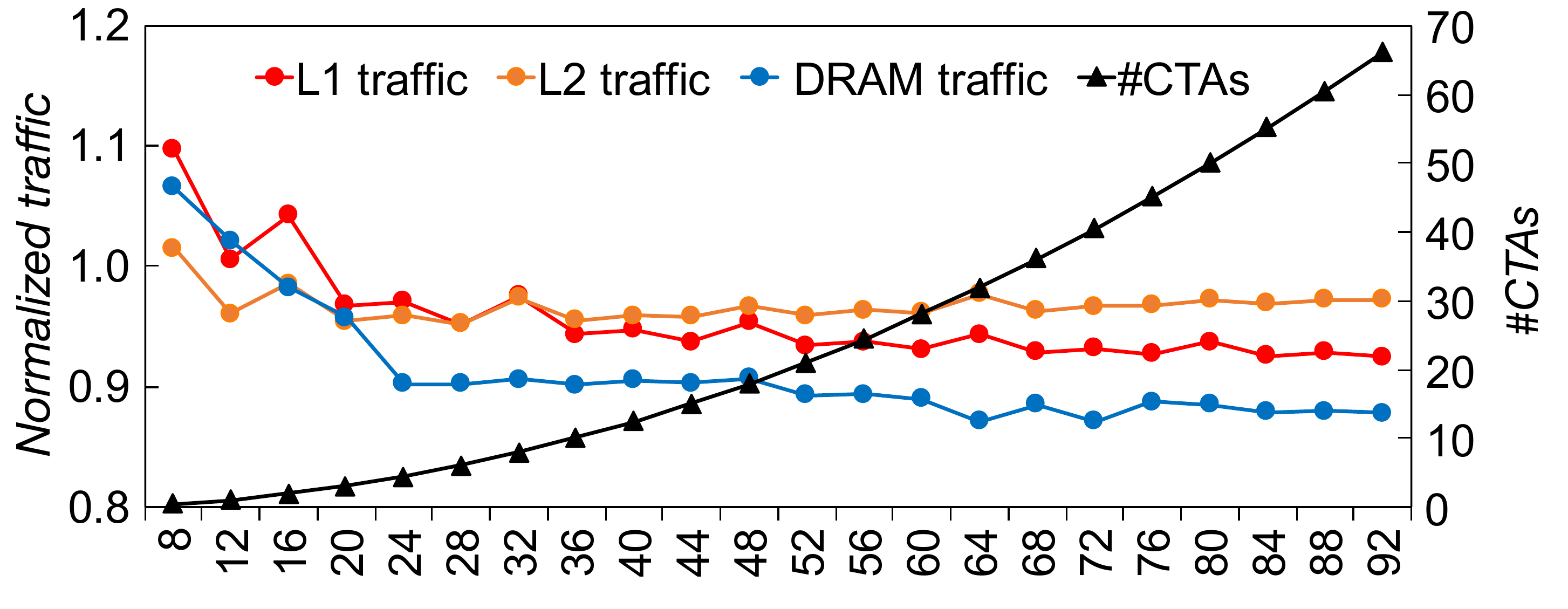}
        \label{fig:sensitivity_HW}
        
    }\\
    \centering                                                                            
    \subfloat[Sensitivity to mini-batch size \((=B)\)]{
        \includegraphics[width=0.47\textwidth]{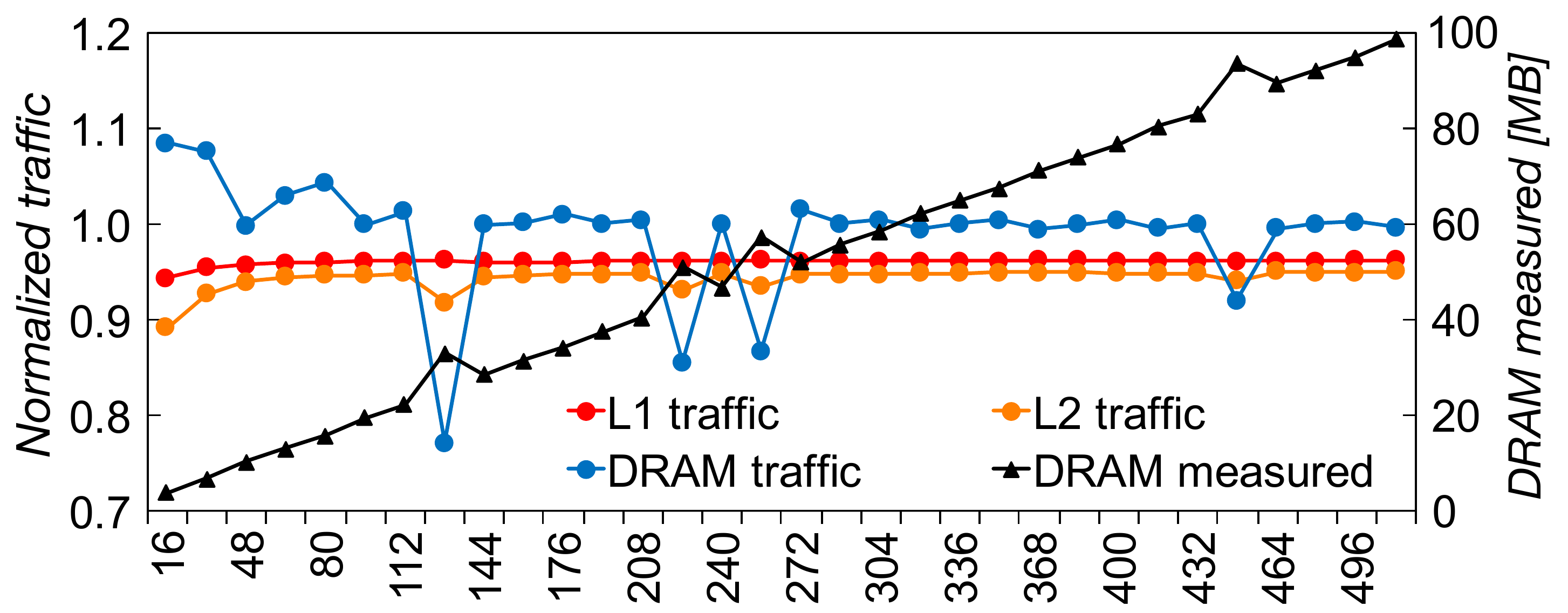}                          
        \label{fig:sensitivity_B}
    }\\
    \caption{\small{Predicted L1, L2, and DRAM traffic by \del using artificial \conv layer configurations (normalized to the measurement).}}
    \label{fig:sensitivity}
    \vspace*{-4mm}
\end{figure}

We analyze the sensitivity of our memory traffic model to artificial \conv layers. We use 256 input channels, a 13$\times$13 IFmap, 128 output channels, a 3$\times$3 filter, and stride 1, which is a commonly-used \conv layer configuration~\cite{he2016deep,szegedy2015going}. Then, we sweep each of these parameters to observe the model's sensitivity. \fig{fig:sensitivity} shows the predicted data traffic at L1, L2, and DRAM channels, which are normalized to the measured data traffic on a TITAN Xp GPU.

\textit{\textbf{Sensitivity to output channel count.}}
For different output channel counts, \del shows geometric mean absolute error (GMAE) of 3.7\%, 8.4\%, and 3.7\% for L1, L2, and DRAM respectively~(\fig{fig:sensitivity_K}). The number of output channels affects the CTA tile width and the error is correlated with this CTA tile width (black line). Based on our profiled data, narrow CTA tiles (32 and 64) use \(blk_K\) size of 4 rather than 8 and this makes one warp access multiple distant filter data columns~(\fig{fig:l1_traffic}), which increases L1 load inefficiency. In case of DRAM traffic, using a narrow CTA tile divides the GEMM matrix into more CTA tiles, increasing the number of CTA columns. As \del calculates DRAM traffic by multiplying the IFmap matrix size and the number of columns of CTA tiles,  data reuse between the columns can increase the error. The L2 estimates show the opposite error trend, which is because the ratio between \(blk_K\) and the filter size is used to calculate \(DIST_H\).

\textit{\textbf{Sensitivity to input channel count.}}
In general, the estimated traffic is not sensitive to the number of input channels~(\fig{fig:sensitivity_C}). Across different input channel counts ranging from 16 to 512, \del shows only 4.1\%, 5.7\%, and 1.2\% GMAE compared to measurements for L1, L2, and DRAM load traffic, respectively. DRAM traffic error has larger variation showing localized drops at some points, which are correlated with local peaks in the measured DRAM traffic (black line).
These are potentially caused by working set alignment to L2 cache associativity that may lead to cache thrashing, which is not considered in our model. Although \del cannot predict such points, its estimated traffic follows the trend across different input channel counts with very small error.

\textit{\textbf{Sensitivity to feature size.}}
\del over-predicts all data traffic of layers with small IFmap sizes (e.g., \(H_i \times W_i <20\))~(\fig{fig:sensitivity_HW}). For L1, decreasing IFmap size increases the range of data that a warp loads and therefore the load inefficiency. Our L1 model slightly overestimates the L1 traffic when the IFmap size is small. For large IFmaps, the predicted L1, L2, and DRAM traffic exhibit small error.

\textit{\textbf{Sensitivity to mini-batch size.}}
The mini-batch size does not affect the im2col GEMM data layout. Therefore, both L1 and L2 traffic prediction results have very small variation for different mini-batch sizes ranging 16 to 512~\fig{fig:sensitivity_B}. When the memory footprint size of the workload is close to the L2 cache capacity, DRAM traffic is slightly overestimated and deviates by up to 10\% compared to measurements. We again observe localized drops in accuracy at some mini-batch sizes (maximum 20\%). We believe this is caused by L2 cache thrashing, as explained earlier.

\renewcommand{\thesubsection}{\Alph{subsection}}

\begin{figure}[h]
    \centering
    \subfloat[Pascal TITAN Xp]{
        \includegraphics[width=0.46\textwidth]{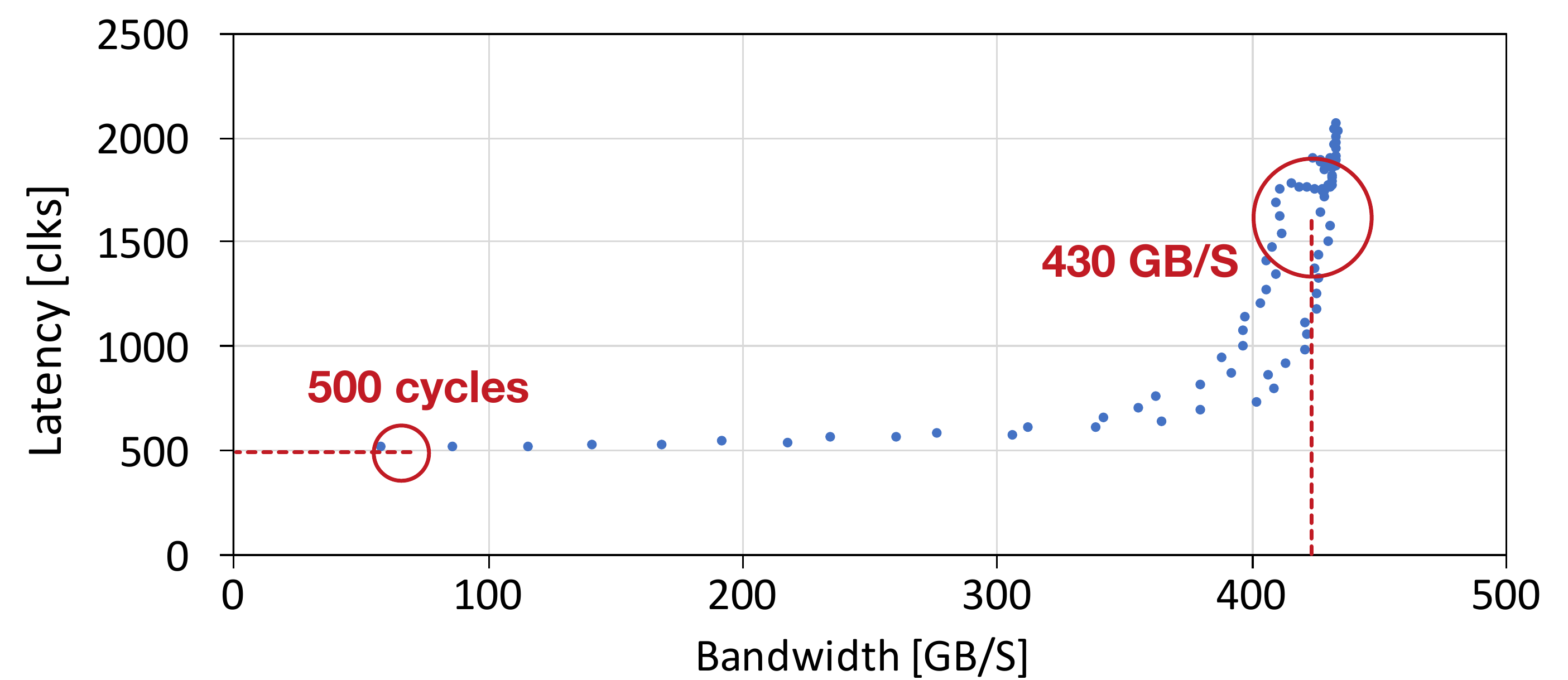}
    }\\
    \vspace*{-3mm}
    \centering
    \subfloat[Pascal TESLA P100]{
        \includegraphics[width=0.46\textwidth]{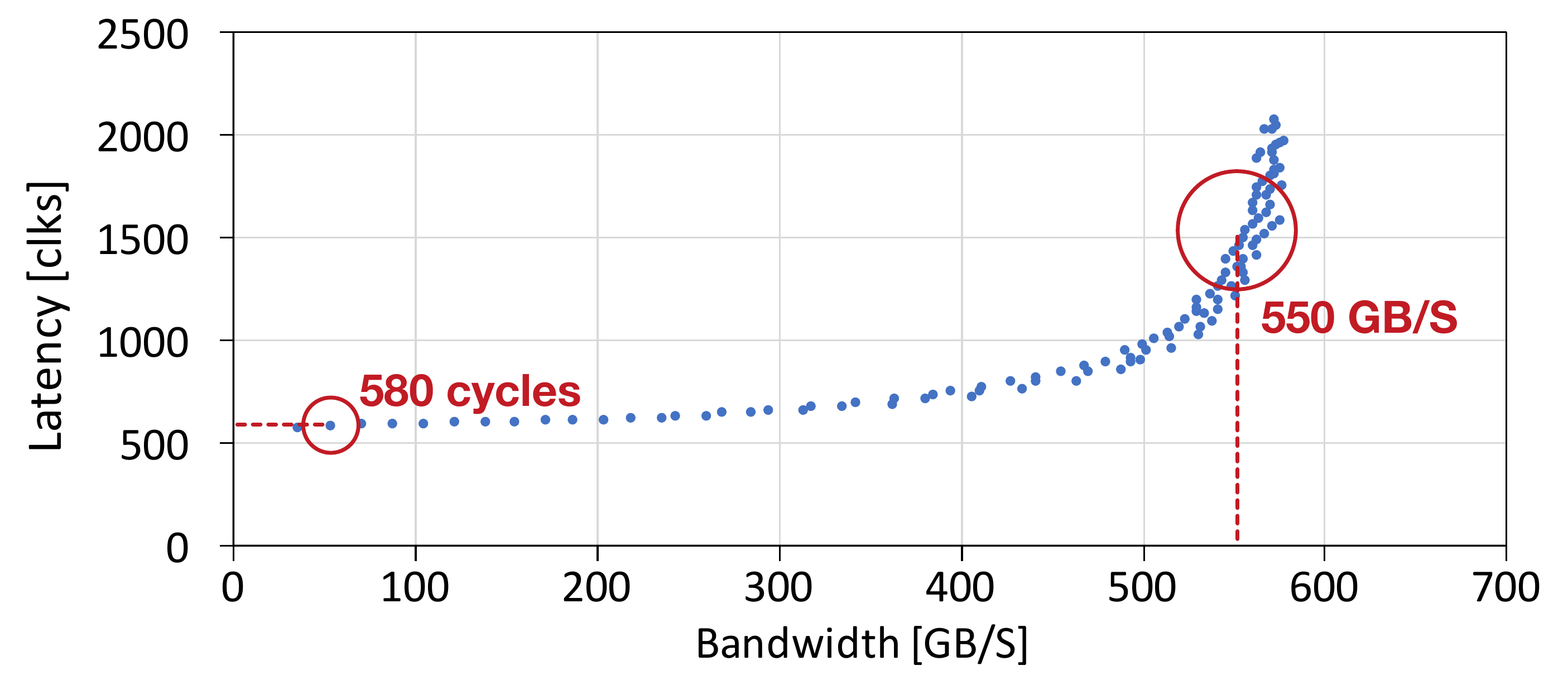}
    }\\
    \vspace*{-3mm}
    \centering
    \subfloat[Volta TESLA V100]{
        \includegraphics[width=0.46\textwidth]{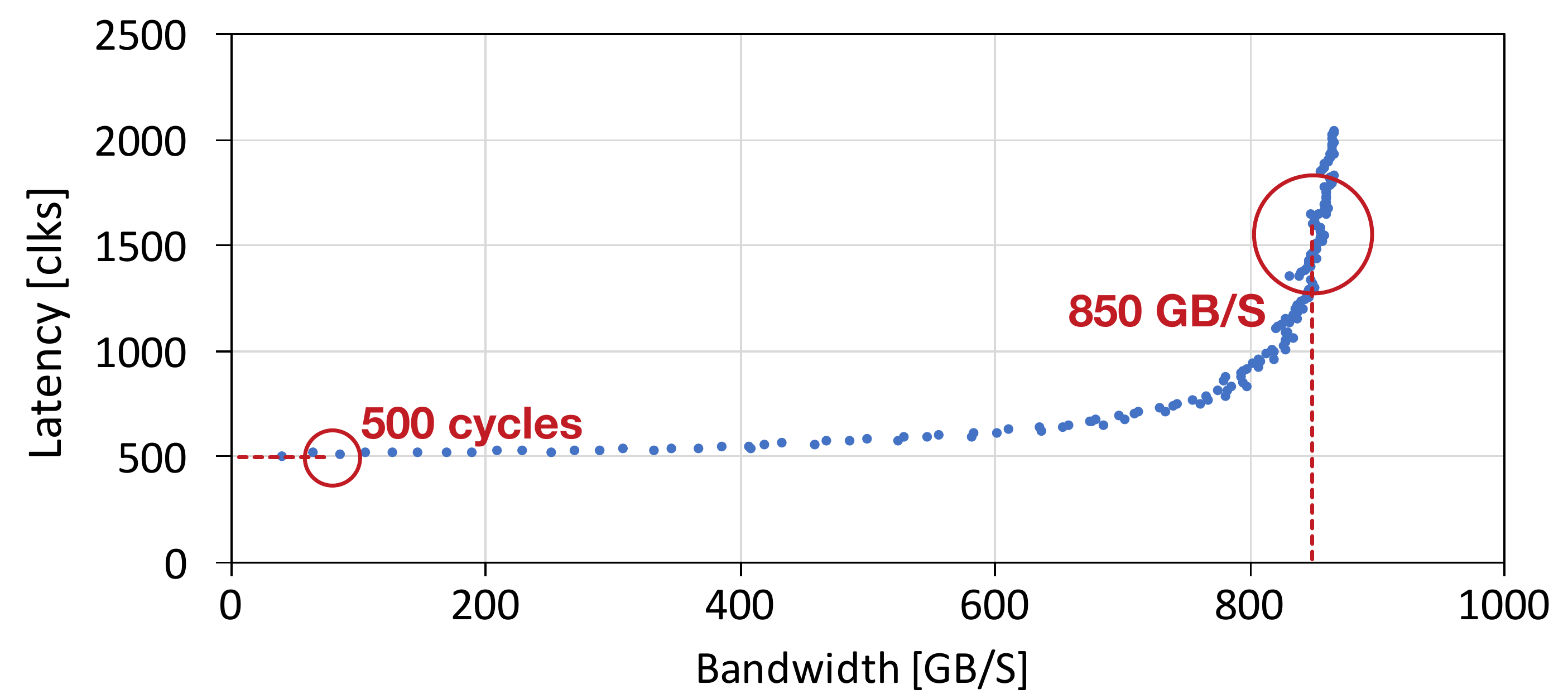}
    }\\
    \caption{\small{DRAM latency and effective DRAM bandwidth of different GPU devices measured using a micro benchmark.}}
    \label{fig:micro_bench}
    \vspace*{-3mm}
\end{figure}

\subsection{DRAM Latency and Effective Bandwidth Measurement}
\label{subsec:micro_bench}
\fig{fig:micro_bench} shows the turnaround latency to DRAM and the effective bandwidth at DRAM channels, which we measure on three GPU devices (TITAN Xp, P100, and V100) used for our experiment. We use a micro benchmark which generates a stream of DRAM traffic with increasing volume per time unit. When the DRAM traffic is small, the measured DRAM latency does not change with increasing memory access intensity. This is because the DRAM data channel is not busy and there is no or minor DRAM bank conflicts. The DRAM latency measured in this range is what we defined as physical latency in \sect{sec:perf_model}. As the memory traffic intensity increases further, the measured DRAM turnaround latency exponentially increases. At this point, the large amount of traffic bottlenecks the DRAM data bus thus the transactions are stuck in the queue. We use the DRAM bandwidth measured for the peak DRAM channel bandwidth. Due to DRAM bank interleaving conflicts, the measured effective DRAM bandwidth is smaller than the theoretical maximum bandwidth of the GPU devices.

\subsection{Execution Cycle Comparison}
\label{subsec:execution_cycles}
\begin{figure}[h]
    \centering
    \subfloat[AlexNet]{
        \includegraphics[width=0.48\textwidth]{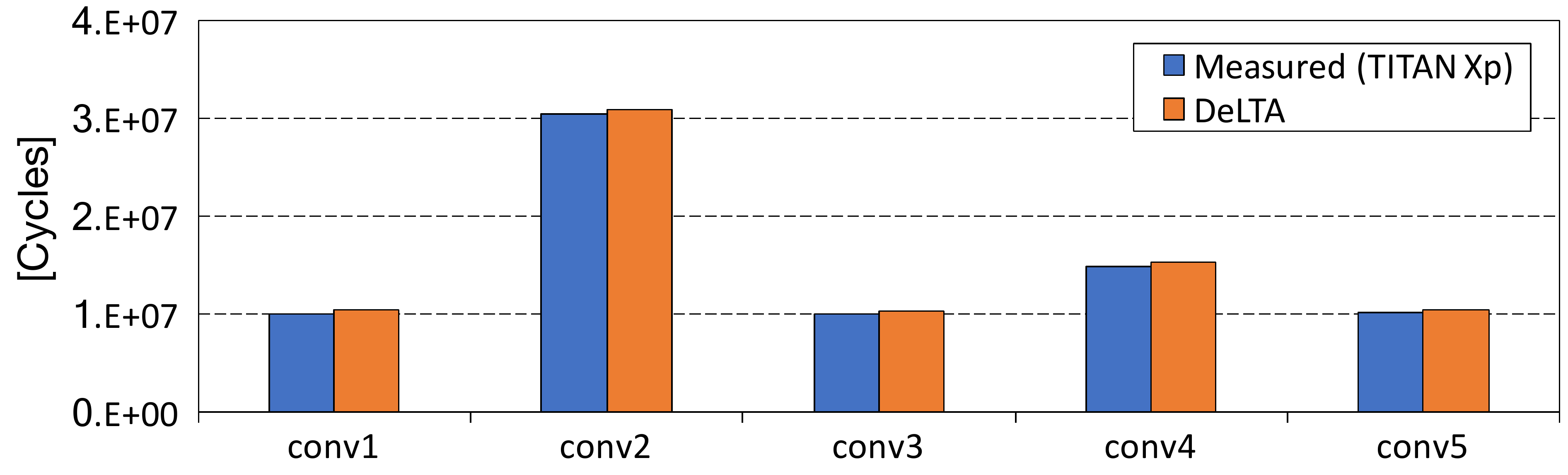}
    }\\
    \vspace*{-3mm}
    \centering                                                                            
    \subfloat[VGG16]{
        \includegraphics[width=0.48\textwidth]{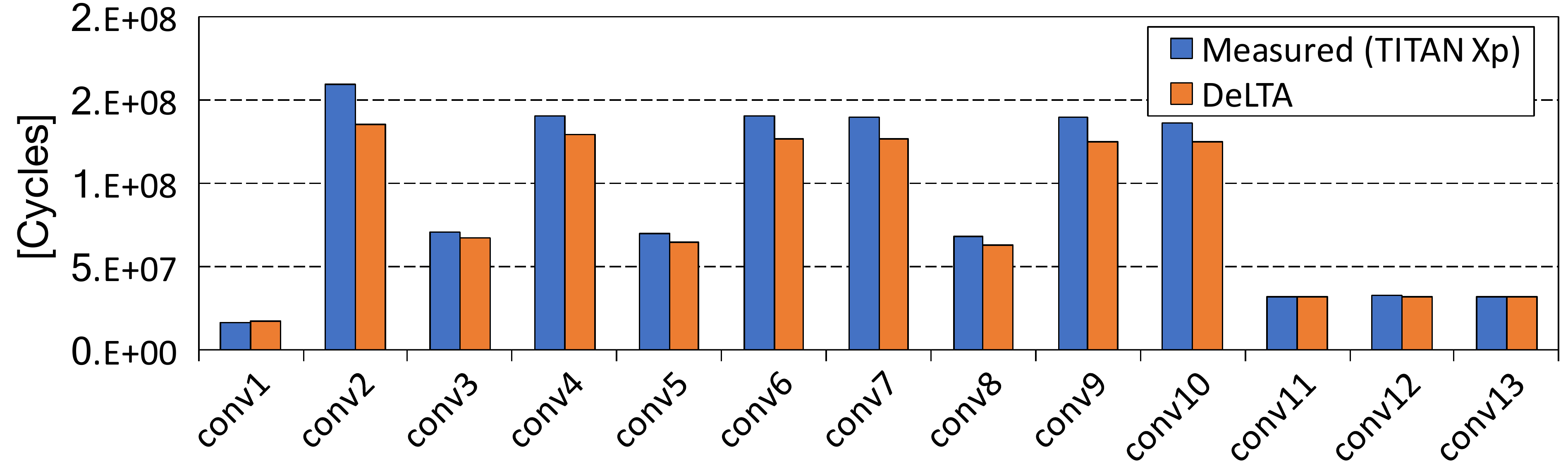}
    }\\
    \vspace*{-3mm}
    \centering
    \subfloat[GoogLeNet]{
        \includegraphics[width=0.48\textwidth]{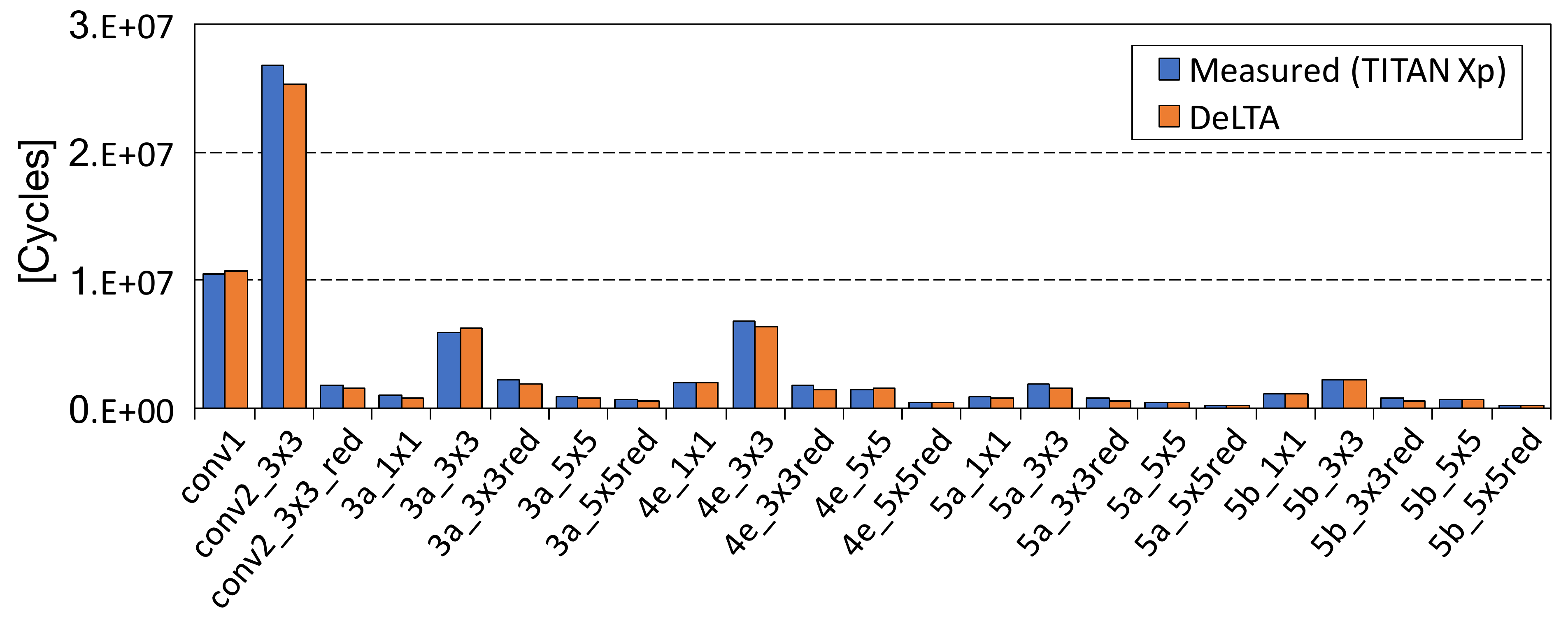}
    }\\
    \vspace*{-3mm}
    \centering                                                                            
    \subfloat[ResNet]{
        \includegraphics[width=0.48\textwidth]{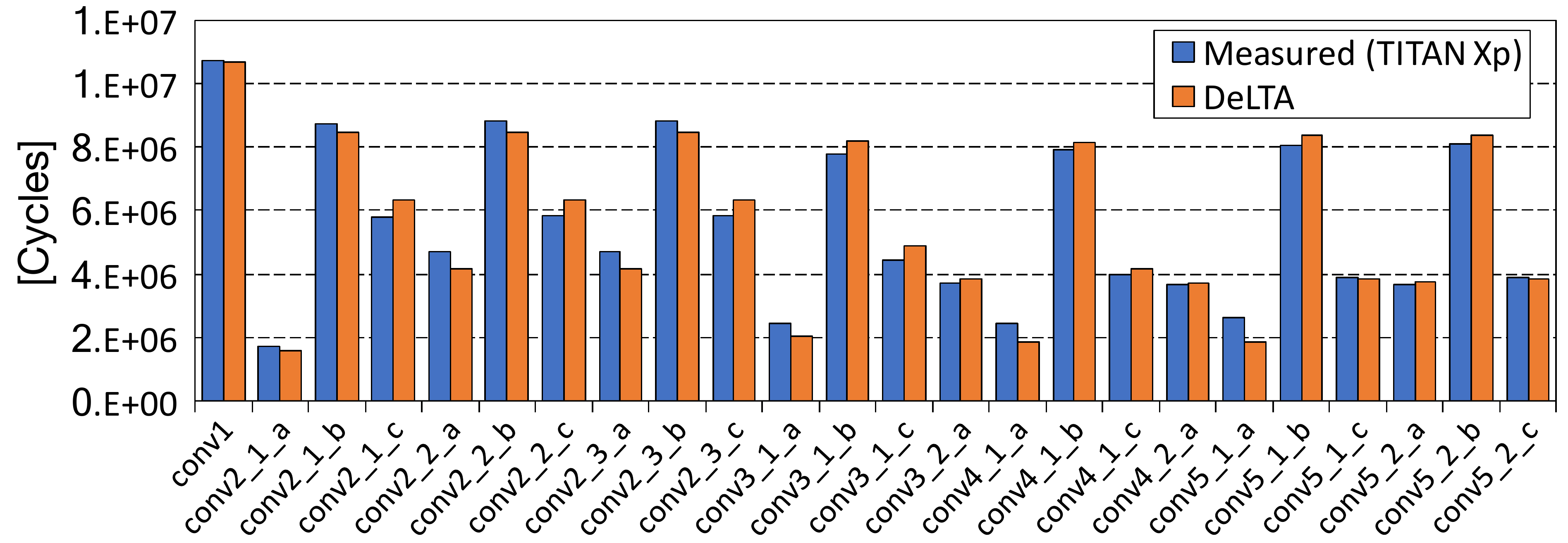}
    }\\
    \vspace*{-2mm}
    \caption{\small{The estimated execution cycles by \del and the measurement on TITAN Xp.}}
    \label{fig:cycle_prediction}
\end{figure}

\fig{fig:cycle_prediction} compares the estimated execution cycles of the conv layers used in four modern CNNs to the measured cycles on Pascal TITAN Xp. The execution cycles are different by an other of magnitude depending on the convolution configuration, and \del shows high estimation accuracy regardless of the total execution cycles.

\subsection{Memory Traffic Comparison}
\label{subsec:memory_traffic}
\begin{figure*}[h]
    \centering
    \includegraphics[width=1.\textwidth]{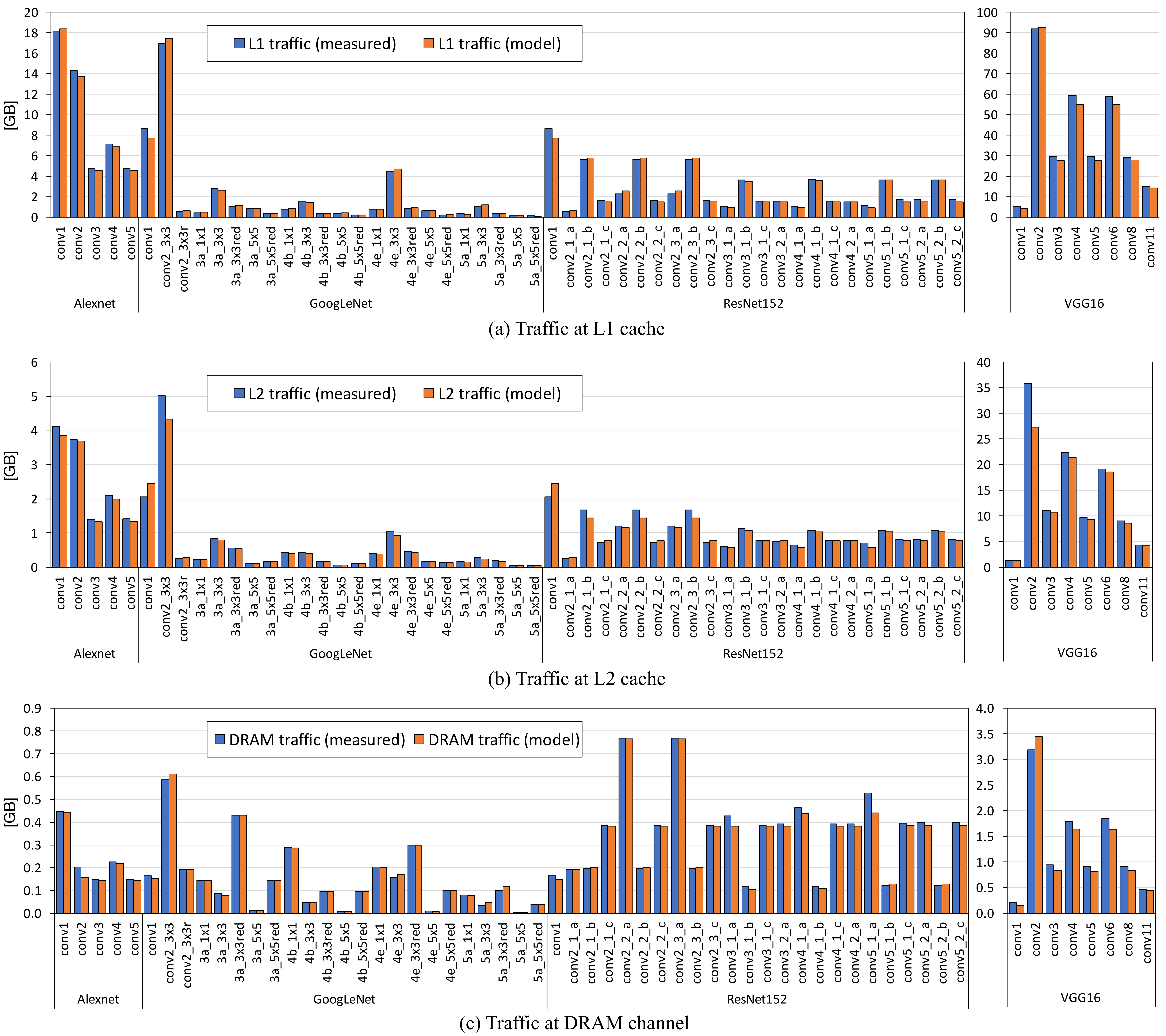}
    \caption{\small{Comparison of the model estimated L1, L2, and DRAM traffic to the measured data on NVIDIA TITAN Xp GPU.}}
    \label{fig:memory_traffic}
\end{figure*}

\fig{fig:memory_traffic} compares the memory traffic at L1, L2, and DRAM channels that are estimated using our model and measured on NVIDIA TITAN Xp GPU. \del shows highly accurate memory traffic prediction regardless of convolution kernel's total memory traffic volume.

\clearpage           

\end{document}